\documentclass[a4paper,fleqn,usenatbib]{mnras}

\usepackage[T1]{fontenc}
\usepackage{ae,aecompl}

\usepackage{graphicx}	
\usepackage{amsmath}	
\usepackage{amssymb}	
\usepackage[autostyle]{csquotes}

\usepackage{xspace}
\xspaceaddexceptions{]\}}
\xspaceaddexceptions{E}
\xspaceaddexceptions{s}

\title[Non-circular flows in HighMass galaxies]{Non-circular flows in HIghMass galaxies in a test of the late accretion hypothesis}

\author[D. Bisaria et. al.]{
Dhruv Bisaria$^{1}$\thanks{E-mail: dhruv.bisaria@queensu.ca},
Kristine Spekkens$^{1,2}$,
Shan Huang$^{3}$,
Gregory Hallenbeck$^{4}$,\newauthor
Martha P. Haynes$^{5}$
\\
$^{1}$Department of Physics, Engineering Physics, and Astronomy, Queen's University, Kingston, ON K7L 3N6, Canada\\
$^{2}$Department of Physics and Space Science Royal Military College of Canada P.O. Box 17000, Station Forces Kingston, ON K7K 7B4, Canada\\
$^{3}$Center for Cosmology and Particle Physics, New York University, New York, NY, 10003, United States\\
$^{4}$Department of Computing and Information Studies, Washington \& Jefferson College, Washington, PA, 15301, United States\\
$^{5}$Cornell Center for Astrophysics and Planetary Science, Cornell University, Ithaca, NY, 14853, United States\\
}

\date{Accepted XXX. Received YYY; in original form ZZZ}

\pubyear{2021}

\begin{document}
\label{firstpage}
\pagerange{\pageref{firstpage}--\pageref{lastpage}}
\maketitle

\begin{abstract}
	
\noindent We present H$\alpha$ velocity maps for the HIghMass galaxies UGC~7899, UGC~8475, UGC~9037 and UGC~9334, obtained with the SITELLE Imaging Fourier Transform Spectrometer on the Canada-France-Hawaii Telescope, to search for kinematic signatures of late gas accretion to explain their large atomic gas reservoirs. The maps for UGC~7899, UGC~9037, and UGC~9334 are amenable to disc-wide radial flow searches with the DiskFit algorithm, and those for UGC~7899 and UGC~9037 are also amenable to inner-disk kinematic analyses. We find no evidence for outer disc radial flows down to $\bar{V}_r \sim 20 \ \kms$ in UGC~9037 and UGC~9334, but hints of such flows in UGC~7899. Conversely, we find clear signatures of inner ($r \lesssim 5$~kpc) non-circularities in UGC~7899 and UGC~9037 that can be modelled as either bisymmetric (which could be produced by a bar) or radial flows. Comparing these models to the structure implied by photometric disc-bulge-bar decompositions, we favour inner radial flows in UGC~7899 and an inner bar in UGC~9037. With hints of outer disc radial flows and an outer disc warp, UGC~7899 is the best candidate for late accretion among the galaxies examined, but additional modelling is required to disentangle potential degeneracies between these signatures in \Hone and H$\alpha$ velocity maps. Our search provides only weak constraints on hot-mode accretion models that could explain the unusually high \Hone content of HIghMass galaxies.

\end{abstract}



\section{Introduction}\label{sec:intro}

Galaxies within halos are surrounded by large coronas of hot gas and grow as low-metallicity gas from the corona and circumgalactic medium accrete onto them \citep{fraternali2017gas, ho2017quasars}. Star-forming galaxies at low redshift have the highest rates of star formation at early times when their gas reservoirs are plentiful. The star formation rate of a typical galaxy decreases as its stellar mass increases: gas within the galaxy is consumed to form stars. For this reason, gas fraction and stellar mass have a strong negative correlation \citep{catinella2010galex,huang2014highmass}. This trend is consistent with a cosmic downsizing scenario whereby as time passes, the galaxies which contribute to the bulk of star formation within the universe get less and less massive \citep{cowie1997evolution,juneau2005cosmic,fontanot2009many}. 

\begin{figure}
	\centering 
	\includegraphics[scale=1.25]{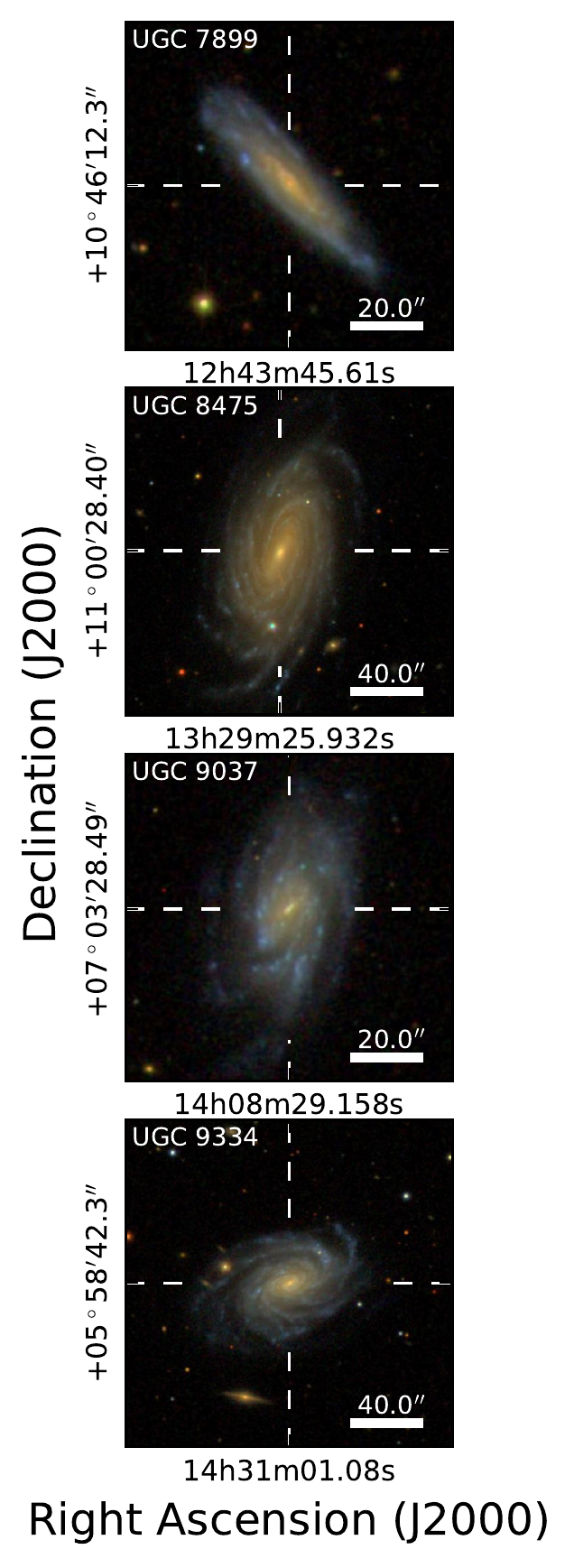}
	\caption[SDSS Optical Images of UGC~7899, UGC~8475, UGC~9037, and UGC~9334]{From top to bottom, SDSS DR14 \citep{Blanton2017} composite images of UGC~7899, UGC~8475, and UGC~9037, and UGC~9334. The white bar in the lower-right corner of each panel shows the angular scale of each image. See Table~\ref{table:observationdetails} for conversions between angular and physical distance.}
	\label{fig:quadimageSDSSDR14}
\end{figure}

The recently completed Arecibo Legacy Fast ALFA (ALFALFA) extragalactic survey detected 34 spiral galaxies that are both massive in atomic hydrogen, \Hone, (\MHI~$>$~10$^{\textrm{10}} \ $M$_{\odot}$) and that have high gas fractions (\MHI/\Mstellar) relative to their stellar masses \citep{huang2014highmass,haynes2018arecibo}. Branded HIghMass, \Hone fractions in the sample range $0.24 \ \lessthan \ \text{\MHI/\Mstellar} \ \lessthan \ 9.2$, with a median gas fraction of $1.23$ \citep{huang2014highmass}. The combination of both high \MHI and high $\MHI/\Mstellar$ distinguish HIghMass from other gas-rich samples such as the ``\Hone Monsters" \citep{lee2014molecular} and ``\Hone GHz" \citep{catinella2014highz} samples. The four HIghMass galaxies studied in this paper are UGC~7899, UGC~8475, UGC~9037, and UGC~9334, and their optical SDSS images are shown in Figure~\ref{fig:quadimageSDSSDR14}. 

Two possible explanations for the overabundance of \Hone in HIghMass galaxies are that they have undergone inefficient star formation over cosmic time, or they have recently accreted external gas to replenish what they depleted to form stars. In other words, they may be either `star-poor' and/or `gas-rich' respectively.

Some HIghMass galaxies may be rendered star-poor if they are embedded in dark matter halos characterized by a high spin parameter ($\lambda$) \citep{peebles1971rotation}, which implies that the galactic disc is extended and sparse \citep{mo1998formation}. Previous studies suggest that the $\lambda$ of some HIghMass galaxies are higher than average when compared to other local galaxies, but that for others is commensurate with the gas-rich galaxy population \citep{hallenbeck2016highmass}. In addition, UV, optical broadband, and \Halpha imaging of HIghMass galaxies \citep{huang2014highmass} suggests that the majority have higher current SFRs than the parent ALFALFA sample, which does not support the high-spin hypothesis.

However, the \Hone morphology and kinematics of HIghMass galaxies provide hints that they may have been inactive in the past. \cite{hallenbeck2014highmass} found evidence for inflowing \Hone across the entire disc of UGC~9037, and \cite{hallenbeck2016highmass} identified a potential \Hone disc warp at $r > 20$~kpc in UGC~7899. In addition, a variety of axisymmetric instability indicators in \Hone and CO \citep{toomre1964gravitational,hunter1998relationship,schaye2004star,romeo2011effective,romeo2013simple} suggest moderate to strongly unstable inner discs in both UGC~7899 and UGC~9037, which may also be consistent with a fresh gas reservoir on the verge of forming stars. These results may suggest that both galaxies have accreted material from external reservoirs or acquired it in a merger event(s).

Under the late accretion scenario, HIghMass galaxies maintain their gas richness through the accretion of low-metallicity gas from the corona and circumgalactic medium. This process occurs after galaxy formation and is hence referred to as late accretion, and may explain numerous phenomena not exclusive to the HIghMass sample \citep{armillotta2016efficiency, fraternali2017gas}. Some degree of accretion must be ongoing to maintain star formation in galaxies \citep{pitts1989chemical, fraternali2012estimating, saintonge2013validation}. Unfortunately, in the absence of detecting accretion directly, we only have circumstantial evidence for its occurence \citep{sancisi2008cold, fraternali2017gas}.

Cosmological simulations predict that galaxies build up their gas reservoirs through either hot-mode accretion, where infalling gas shock heats to the halo virial temperature before condensing onto the disc, or cold-mode accretion, in which infalling gas at temperatures lower than virial flows directly into the galaxy \citep{keres2005galaxies}. Analytic models of outside-in hot-mode accretion presented in \cite{pezzulli2016accretion} imply that radial flows for a Milky Way like disc are small over cosmic time ($\lesssim1 \ \kms$), while cold-mode accretion is most common in low-mass systems (total baryon mass $M_{\text{gal}} \ \lesssim 10^{10.3} \ \Msolar$) and dominant at high redshift ($z > 3$). However, the unusually large gas reservoirs of the HIghMass galaxies combined with hints of disc-wide non-circular flows in \Hone make them good candidates in which to search for kinematic signatures of cold or hot accreting gas.

As such, searching for non-circular flows in \Halpha, beginning with a disc-wide search for radial flows, is a logical next step in this analysis, since they can have larger amplitudes in H$\alpha$ than in \Hone \citep{sellwood2010quantifying}, and because \Halpha maps galaxies at higher angular resolution, especially in the inner disc where \Htwo regions are more abundant. 

This paper presents a search for signatures of late gas accretion in four HIghMass galaxies by characterizing non-circular flows in their \Halpha velocity maps. We describe observations of UGC~7899, UGC~8475, UGC~9037, and UGC~9334 with the SITELLE Imaging Fourier Transform Spectrometer (IFTS) at the Canada-France-Hawaii Telescope (CFHT) in Section~\ref{sec:observations_datareduction}. We search the resulting \Halpha velocity maps for radial and bisymmetric non-circular flows using the DiskFit algorithm \citep{spekkens2007modeling,sellwood2010quantifying,sellwood2015diskfit} in Section~\ref{sec:dataanalysis}, and discuss our results in context of the late accretion hypothesis in Section~\ref{sec:discussion}.

\section{Observations and Data Reduction}\label{sec:observations_datareduction}

\subsection{Sample selection}\label{subsec:sampleselection}

We select galaxies from the HIghMass sample defined in \cite{huang2014highmass} that are suitable for our study. Eligible galaxies with significant \Halpha emission need to have intermediate inclinations (45$^\circ \leq i \leq $ 75$^\circ$): galaxies with high inclinations are not amenable to searches for non-circular flows, while those with low inclinations will exhibit little Doppler shift across the disc. In addition, the \Halpha emission of the targets must fall entirely within SITELLE's $647 - 685$~nm SN3 band \citep{drissen2019sitelle}, which is likely if their \Hone-measured systemic velocities satisfy $V_\text{\Hone} \lesssim 13000 \ \kms$. From the 10 HIghMass galaxies that fit these criteria \citep{huang2014highmass}, we were able to observe UGC~7899, UGC~8475, UGC~9037 and UGC~9334.

\subsection{SITELLE Observations} \label{subsec:SITELLEobservations}

UGC~7899 and UGC~9037 were observed during photometric conditions in queue observing mode with SITELLE on 2016 March 6, while UGC~8475 and UGC~9334 were observed on 2017 July 3rd and 2019 April 5th respectively. SITELLE is an Imaging Fourier Transform Spectrometer (IFTS), which produces spectra for each pixel in its field of view \citep{drissen2019sitelle}. The observations were configured to detect \Halpha with a resolution $R \sim 1500$ in the SN3 filter. The corresponding number of $\Delta \lambda$~=~0.45 nm moveable mirror steps and reference wavelength for each observation are given in Table \ref{table:observationdetails}. The spectra contained the  [N II]~6548~\angstrom and 6583~\angstrom doublet, in addition to the brighter \Halpha line, which was fit along with \Halpha to extract line-of-sight velocities. Because the N II doublet is weaker than the \Halpha emission line, the effect on the fits is mild and we do not consider the N II maps in this analysis \citep{drissen2019sitelle}.

Table \ref{table:observationdetails} also contains the Moon's illumination fraction f$_{\text{illum}}$ and its mean angular separation $\bar{\theta}_{\text{sep}}$ from the target during the observations. Distributed photon noise from the Moon limits the sensitivity of SITELLE \citep{drissen2014imaging}, making dark sky conditions ideal. While this was the case for the UGC~7899, UGC~9037, and UGC~9334 runs, the Moon phase and position were more problematic for our run on UGC~8475. We compensated for the higher lunar photon noise by increasing the integration time per step on this target. Even so, the resulting data quality is lower for this galaxy (see Section~\ref{subsec:finalmaps}).

\begin{table*}
	\centering
	\begin{tabular}{c|cc|ccccc|ccc} \hline \hline
		Galaxy & $\Phi$ & $d$ & $D$ & $N$ & t$_{\text{exp}}$  &  $\lambda_{\text{ref}}$  & $\bar{\theta}_{\text{sep}}$ & $f_{\text{illum}}$ & \vthreshold & \Fthreshold  \\
		~ & [\arcsec] & [Mpc] & [$\frac{\text{kpc}}{\arcsec}$] & ~ & [s]  &  [nm]  & [\degreesign] & ~ & [\kms] & [\fluxunits]  \\
		(1) & (2) & (3) & (4) & (5) & (6) & (7) & (8) & (9) & (10) & (11) \\ \hline
		\multicolumn{1}{r}{UGC 7899} & 0.32 & 129 & 0.63 & 264   &  22.0  & 675.9  & 125 & 0.08 & 16  & $\dots$  \\
		\multicolumn{1}{r}{UGC 8475} & 0.65 & 102 & 0.49 & 206 & 56.2 & 671.6  & 34 & 0.80 & 15 & $4.8 \times 10^{-17}$  \\
		\multicolumn{1}{r}{UGC 9037} & 0.32 & 88 & 0.43 & 259 & 22.0 & 669.5 & 78 & 0.24 & 15 & $\dots$  \\
		\multicolumn{1}{r}{UGC 9334} & 0.32 & 110 & 0.53 & 260 & 25.0 & 673.1 & 148 & 0.01 & 15 & $\dots$  \\
		\hline
	\end{tabular}
	\caption[Observation Details]{Galaxy and observation parameters. Cols 2, 3 and 4: the angular resolution per pixel, the distance to the galaxy in megaparsecs from the complete ALFALFA catalogue \citep{haynes2018arecibo}, and the resolution at that distance, respectively. Col 5: number of steps of the moveable mirror in the IFTS (and thus, the number of interferometric images taken). Col 6: the exposure time per step. Col 7: reference wavelength, $\lambda_{\text{ref}}$, at which $R = \lambda_{\text{ref}}/\Delta \lambda$ = 1500, where $\Delta \lambda$ is the step size. Cols 8 and 9: the lunar separation and the lunar illumination fraction respectively during observations. Col 10: the velocity-uncertainty threshold value. Pixels below this value are included in the filtered velocity map. Col 11: the flux threshold, above which pixels are included in the velocity map. Only necessary for UGC~8475.}
	\label{table:observationdetails}
\end{table*}

\subsection{Data Reduction with ORCS}\label{subsec:dataredORCS}

\begin{figure}
	\centering
	\includegraphics[scale=0.8]{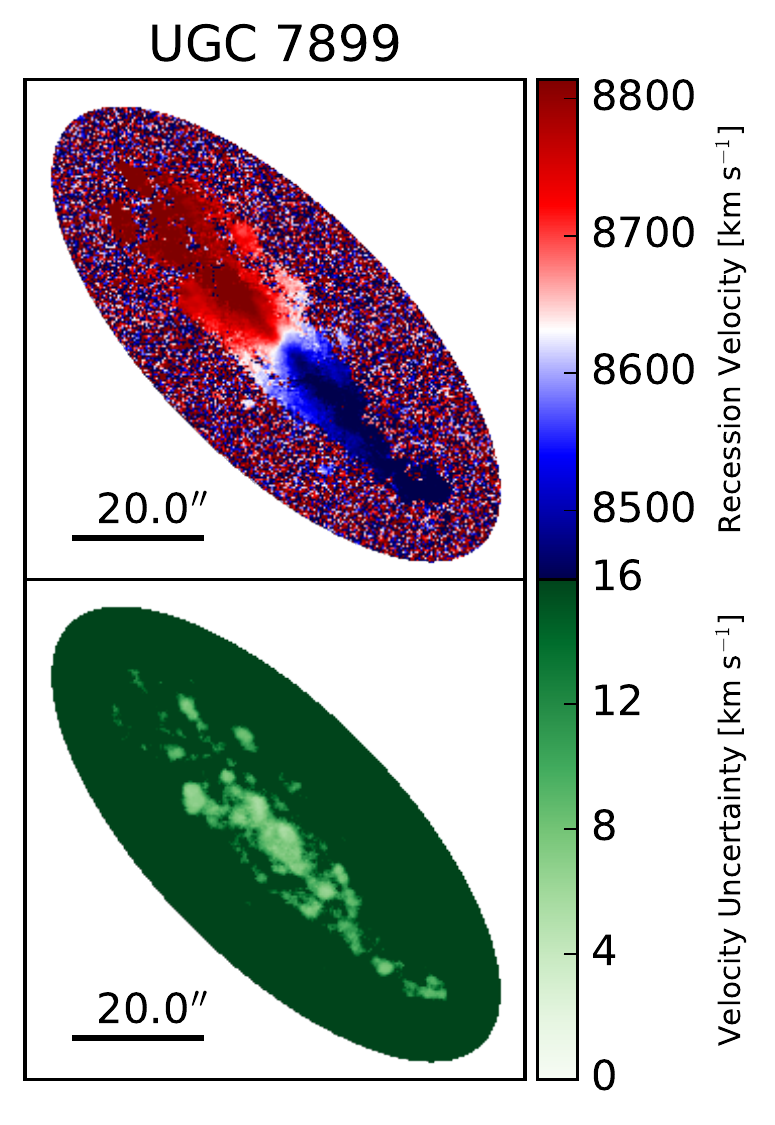}
	\caption[UGC 7899 Composite Velocity and Uncertainty Map]{The velocity (top) and velocity-uncertainty (bottom) maps of UGC~7899 output from the ORCS velocity field script described in Section~\ref{subsec:dataredORCS}. The black bar in the lower-left corner shows the angular scale of each panel, and the ellipse containing non-zero data is the ROI for this galaxy. The maximum value of the colour bar corresponds to \vthreshold for UGC~7899.}
	\label{fig:verticalreganderrUGC7899}
\end{figure}

ORCS is a publicly available\footnote{http://celeste.phy.ulaval.ca/orcs-doc/index.html} analysis engine for spectral data cubes taken by SITELLE \citep{martin2015orbs}. ORCS generates velocity maps from hyperspectral cubes by fitting an emission line profile (sinc or sincgauss, the convolution of a sinc and a Gaussian) to the spectrum of each pixel in a user-generated Region of Interest (ROI) around the galaxy within SITELLE's field of view. The fitting algorithm requires an initial guess for the emission line velocity, which should be no more than a FWHM from the emission line profile centroid to return a reliable fit. For a sinc profile, this is equal to FWHM $= 1.21\Delta w$, where $\Delta w$ is the velocity spectrum channel width. Since our targets are massive disc galaxies at intermediate inclinations, their spectral extents vastly exceed the FWHM of our observing set-up. For each target we therefore scripted ORCS to perform a series of sinc-profile fits using different input velocities across an ROI encompassing the entire galaxy to create a series of partial maps. We use the velocity-uncertainty associated with a given pixel as a proxy for the goodness of fit to each spectrum. Reliable fits have significantly lower velocity uncertainties than unreliable ones, and we therefore construct a velocity field for each target by selecting the pixel values within the ROI with the lowest velocity uncertainty across all partial maps. We show the complete velocity and velocity uncertainty maps of UGC~7899 that result from this approach as an example in Figure~\ref{fig:verticalreganderrUGC7899}. 

\subsection{Final Velocity Maps}\label{subsec:finalmaps}

Velocity field pixels with the largest uncertainties tend to be those for which ORCS failed to find real emission and instead returned the velocity of a spurious noise peak in the vicinity of an input guess. In the outskirts of the galaxy disc, spurious velocities quickly outnumber reliable ones and destabilize the DiskFit models described in Section~\ref{sec:dataanalysis}. To avoid this, we filter out pixels with a velocity uncertainty larger than a given value \vthreshold from the velocity maps returned by the ORCS script described in Section~\ref{subsec:dataredORCS}. The threshold values, \vthreshold are given in Table \ref{table:observationdetails}. Changes to \vthreshold in the range 15-25 \kms produce minor changes to the filtered velocity field input to DiskFit, but do not change the characteristics of the models obtained. We note that for the noisier UGC~8475 data, we additionally retained only velocity map pixels at locations where the H$\alpha$ flux exceeded $ \Fthreshold = 4.8 \ \times \ 10^{-17} \ \fluxunits$. This is suboptimal, as a low-flux pixel can still give an accurate measure of gas velocity so long as the line peak is measured precisely. We did not bin pixels spatially for UGC~7899, UGC~9037, or UGC~9334, but used a bin size of 2 pixels for UGC~8475 to increase SNR at the cost of angular resolution. We experimented extensively with various bin sizes in order to improve the filling factor of our velocity maps, but found that binning pixels for galaxies other than UGC~8475 had no noticeable impact on the kinematic models.

\begin{figure*}
	\centering
	\includegraphics[scale=0.94]{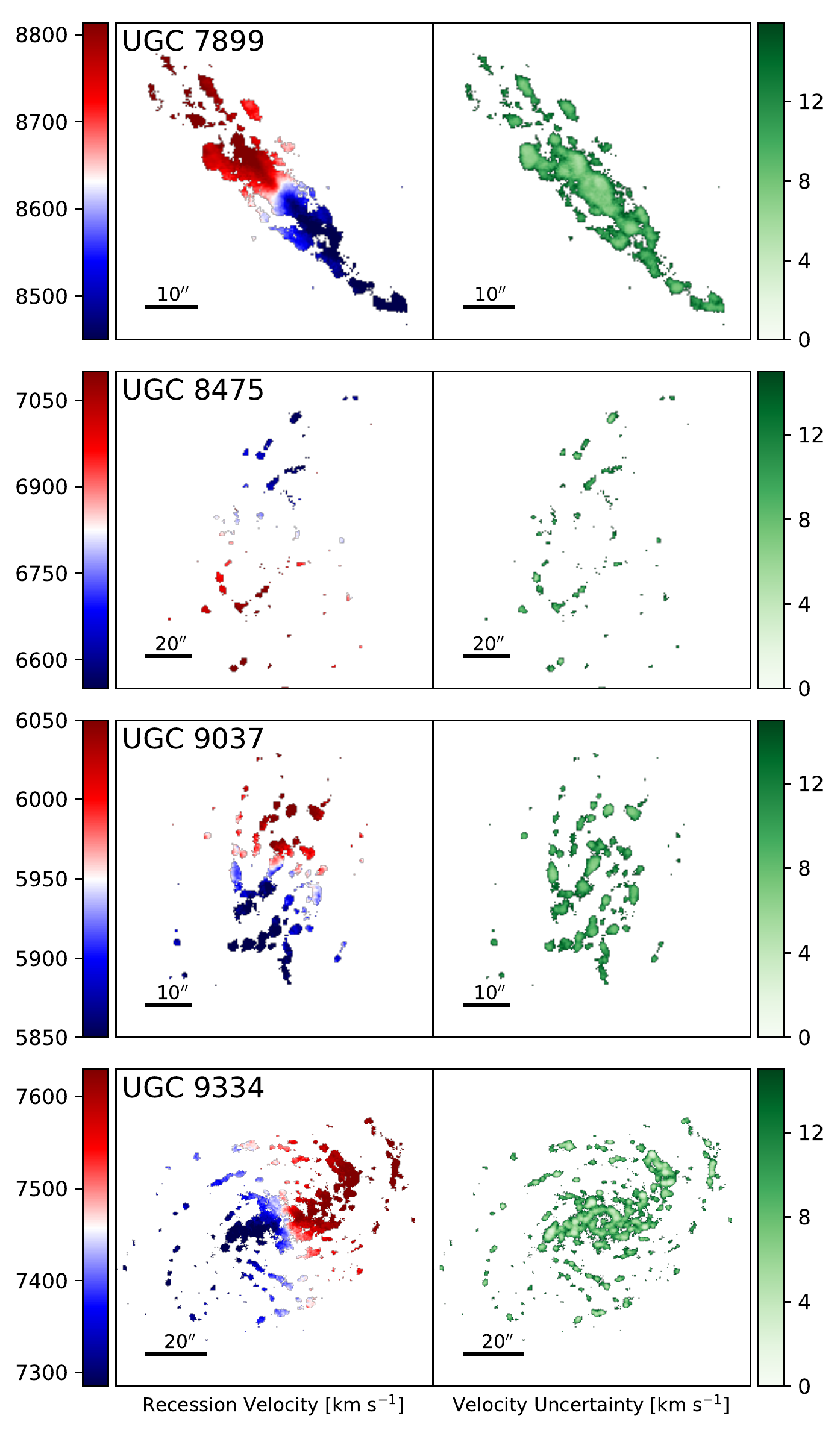}
	\caption[Final Velocity Maps]{From top to bottom, the final velocity and velocity-uncertainty maps of UGC~7899, UGC~8475, UGC~9037, and UGC~9334. Velocity maps are in the left column and velocity-uncertainty maps are in the right column. The black bar in the lower-left corner of each panel shows the angular scale of the maps.}
	\label{fig:filteredvelmaps}
\end{figure*}

Figure~\ref{fig:filteredvelmaps} presents the final velocity and velocity-uncertainty maps for UGC~7899, UGC~8475, UGC~9037, and UGC~9334. Not all galaxy maps are filled out to the same extent. For example, though the observing conditions for UGC~7899 were better than they were for UGC~9037, they were less optimal than the conditions for UGC~9334. In addition, UGC~7899 appears to be more complete relative to UGC~9037 and UGC~9334 because it is at a higher inclination. Despite the differences in data quality, all galaxies show signs of ordered rotation with no obvious signs of distortion, motivating a more quantitative analysis to constrain non-circular flows.

The patchy nature of the velocity maps suggests that the observations trace spiral arms and the \Htwo regions within them, as opposed to the diffuse ionized gas (DIG) (e.g. \citealt{haffner2009warm}). This is especially apparent in UGC~8475, UGC~9037, and UGC~9334, where their moderate inclinations allow for a clearer distinction between arm and interarm regions than in higher-inclination systems such as UGC~7899. Unfortunately, since our observations probe all ionized gas, we cannot disentangle \Htwo regions from the DIG, even though the latter has lower signal-to-noise. As described in Section~\ref{sec:dataanalysis} below, the kinematic models that we apply to the velocity maps are sensitive to coherent non-circular flows rather than local ones associated with individual \Htwo regions or spiral patterns. The velocity maps in Figure~\ref{fig:filteredvelmaps} are therefore amenable to searches for radial flows from late accretion.

\section{Data Analysis with DiskFit}\label{sec:dataanalysis}

In this section we model the SITELLE velocity maps of HIghMass galaxies presented in Figure~\ref{fig:filteredvelmaps} to search for non-circular flows. We describe the kinematic DiskFit models that we adopt in Section~\ref{subsec:kinematicmodels} and the application of those models to the velocity maps in Sections~\ref{subsec:ncflowsalldisk} and \ref{subsec:appofkinmodels}. We combine our kinematic models of UGC~7899 and UGC~9037 with photometric decompositions to determine our preferred models for each of these galaxies in Section~\ref{subsec:noncircflows78999037}. DiskFit is a publicly available code\footnote{\href{https://www.physics.queensu.ca/Astro/people/Kristine\_Spekkens/diskfit/}{DiskFit Source}} developed by \cite{spekkens2007modeling}, \cite{sellwood2010quantifying}, and \cite{sellwood2015diskfit} for the purpose of modelling asymmetries in the photometry (images) or kinematics (velocity maps) of disc galaxies.

\subsection{Kinematic Models}\label{subsec:kinematicmodels}

The kinematic branch of DiskFit applies physically-motivated models to velocity maps of galaxies in order to constrain the disc geometry and kinematic components. While DiskFit has the capacity to fit for simple warps \citep{sellwood2015diskfit}, we consider only models of thin flat discs defined by a kinematic centre (\xcenter,~\ycenter), disc position angle (\diskPA), heliocentric systemic velocity (\Vsys), and an inclination (\diskinc), that do not vary with radius. A thin, flat disc is a reasonable assumption for the HIghMass galaxies given their intermediate inclinations, the lack of strong warps in their \Hone distributions \citep{hallenbeck2016highmass} as well as the fact that \Halpha lies primarily within the optical radius beyond where warps tend to begin. In particular, the radius of the disc at the 25\textsuperscript{th} mag/arcsec\textsuperscript{2} isophote well exceeds the extent of our velocity maps for all of the galaxies \citep{huang2014highmass}, and warps in the radial range probed by our maps are therefore unlikely \citep{briggs1990rules,garcia2002neutral}.

We consider three physically-motivated kinematic models. The rotation-only model assumes that all of the detected gas follows circular orbits as a function of galactocentric radius $r$ about the galaxy centre:

\begin{equation}
\Vmodel = V_{\text{sys}} + \sin(i) \left[  \Vtbar \cos(\theta)\right],
\label{eq:tangentialflowequation}
\end{equation} 

\noindent where \Vmodel is the modelled line of sight velocity, $\bar{V}_{t}(r)$ is the mean tangential velocity component at a given galactocentric radius, and $\theta$ is measured relative to the major axis in the plane of the disc.

We also consider models with radial flows which may be indicative of accreting gas. In the radial flow model, gas flows axisymetrically on both tangential and radial orbits:

\begin{equation} \label{eq:radflowequation}
\Vmodel = V_{\text{sys}} + \sin(i) \left[  \Vtbar \cos(\theta) + \Vrbar \sin(\theta) \right],
\end{equation}

\noindent where \Vrbar is the mean radial velocity component at a given radius. The physical interpretation of \Vrbar depends on which side of the disc minor axis is nearest to the observer: $\Vrbar > 0$ corresponds to inflows if the nearest point of the disc has $\theta = 90\degreesign$, and outflows if the nearest point has $\theta = 270\degreesign$. DiskFit allows non-circular components such as $\bar{V}_{r}$ to be fit across the entire disc or out to a user-specified truncation radius $r_t$. Equations~\ref{eq:tangentialflowequation} and \ref{eq:radflowequation} highlight the potential degeneracy between the disc kinematics and geometry, whereby errors in the latter can produce spurious radial flows or rotation curve features (e.g. \citealt{wong2004search,spekkens2007modeling,sellwood2021uncertainties}).

While the goal of this work is to search for radial flows, bisymmetric flows such as those produced by bars have a similar signature in velocity maps because both $m = 0$ and $m = 2$ harmonics in the disc plane project to $m^{\prime} = 1$ harmonics in the sky plane \citep{schoenmakers1997measuring,spekkens2007modeling}. We therefore also consider a bisymmetric model for non-circular flows detected in the inner disc:

\begin{equation}
\begin{split}
\Vmodel = V_{\text{sys}} + \sin(i) \Big[  & \Vtbar \cos(\theta) - \Vtbarbar \cos(2 \thetab) \cos (\theta) \\ 
- & \Vrbarbar \sin(2\thetab) \sin (\theta) \Big],
\label{eq:m2perturbations}
\end{split}
\end{equation}

\noindent where \Vtbarbar and \Vrbarbar are the tangential and radial components of the bisymmetric flow for $r < r_t$ at a fixed bar position angle \barPA in the plane of the disc, and $\thetab = \theta - \barPA$. Since \phib does not depend on $r$, the bisymmetric model is not sensitive to flows in spiral arms, which DiskFit is not equipped to probe. Equation~\ref{eq:m2perturbations} shows that \Vtbarbar or \Vrbarbar become degenerate with \Vtbar when $\phib = 0$ or $\phib = \frac{\pi}{2}$. This implies DiskFit cannot disentangle bisymmetric flows close to the major or minor axis of the galaxy from rotation (e.g. \citealt{randriamampandry2016exploring}).

The sky-plane bar angle \phibsky of the bisymmetric flow can be calculated from the disc-plane bar angle \phib and the disc geometry \citep{spekkens2007modeling}:

\begin{equation} \label{eq:sky_to_disk_conversion}
	\phibsky = \diskPA + \arctan(\tan\phib \cos i) \text{.} 
\end{equation}

Because of the $2\thetab$ dependence of the non-circular flow terms in Equation~\ref{eq:m2perturbations}, the best fitting \phib can represent either the major or minor axis of the corresponding bar-like distortion to the potential. For bisymmetric flows caused by a stellar bar, the sky plane position angle therefore corresponds to \phibsky from Equation~\ref{eq:sky_to_disk_conversion} computed from either \phib or ($\phib + 90\degreesign$). We report both possibilities for \phibsky for our bisymmetric models.

DiskFit accepts a number of input parameter values. The geometric parameters that were allowed to vary in the models we consider include the disc position angle \diskPA, inclination $i$, best fitting center (\xcenter,~\ycenter) and \Vsys. In the case of a bisymmetric flow model, \phib is also a free parameter. The model also depends on fixed parameters to specify the spatial extent of the fit, such as the radial model extent as well as $r_t$ for any non-circular flows.

In Sections~\ref{subsec:ncflowsalldisk} and \ref{subsec:appofkinmodels}, we describe our application of these kinematic models to the final velocity and velocity-uncertainty maps for UGC~7899, UGC~8475, UGC~9037 and UGC~9334 shown in Figure~\ref{fig:filteredvelmaps}. In all models, parameter uncertainties are determined from 1000 radial-rescaled bootstrap realizations of the velocity maps as described in \cite{sellwood2010quantifying}. In order to assess the validity of features in best-fitting models to each velocity map as a whole, we also model subsets of the data defined by their kinematics. Specifically, we isolate the approaching and receding halves of the disc about the best-fitting minor axis, as well as the Northern and Southern halves above and below the best-fitting major axis. Segmenting was performed using the disc geometry from the best-fitting rotation-only model. When modelling the segmented halves, we let the position angle and inclination vary (but account for the inclination in comparing the rotation curves of different segments) and keep the centre fixed to that obtained by the rotation-only model of the whole galaxy. Finally, we use the distances in Table~\ref{table:observationdetails} to convert galactocentric radius from angular to linear units when plotting kinematic components. 

\subsection{Searching for radial flows across the entire disc}\label{subsec:ncflowsalldisk}

\begin{figure}
	\centering
	\includegraphics[width=0.45\textwidth]{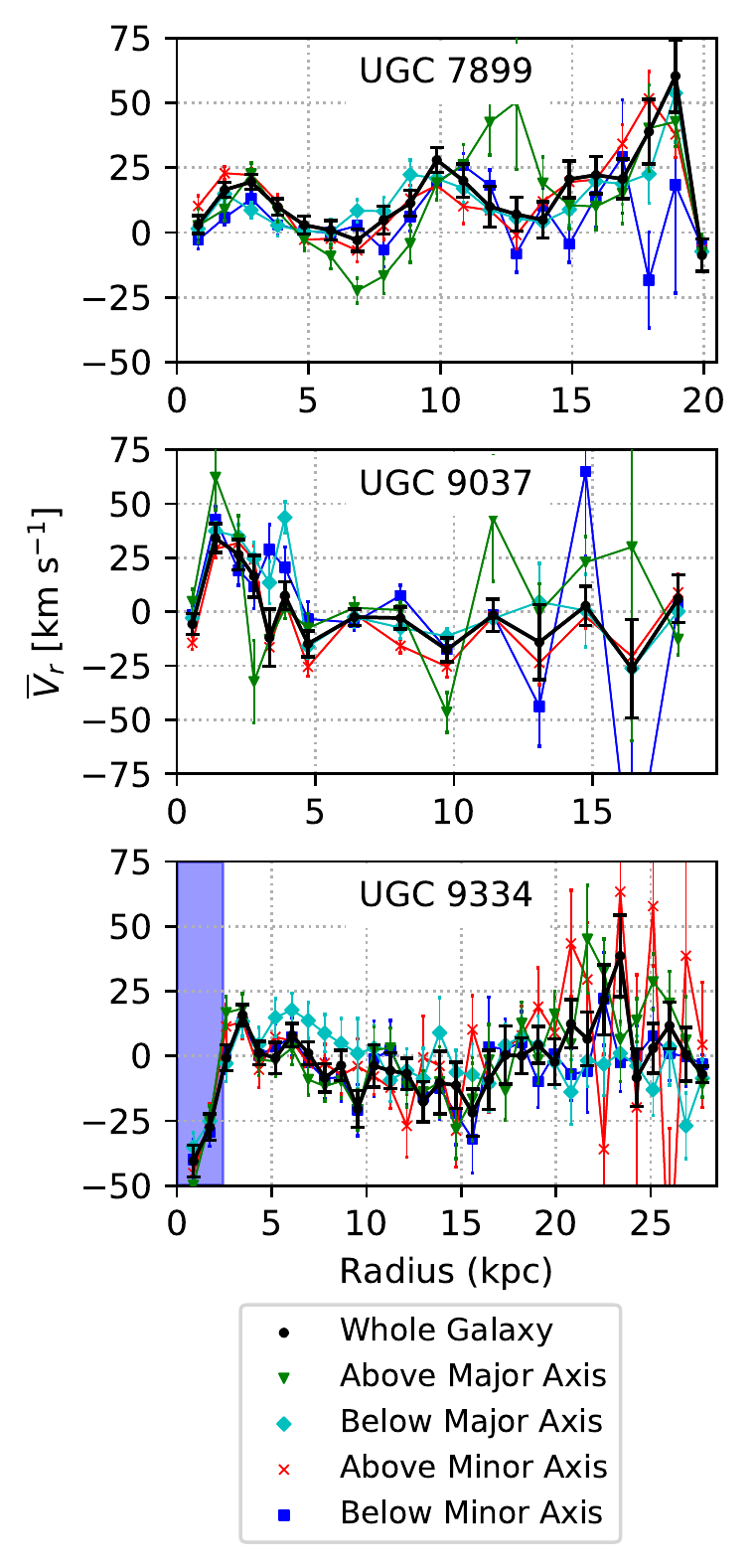}
	\caption[Total-disc radial flows for UGC~7899, UGC~9037, and UGC~9334]{From top to bottom, \Vrbar in radial flow models applied across the entire discs of UGC~7899, UGC~9037, and UGC~9334. In all panels, the black lines show \Vrbar obtained by modelling the velocity field as a whole, whereas the coloured lines represent models applied to subsets of the data. The green upside-down triangles and cyan diamonds show \Vrbar for halves North and South of the major axis. The red crosses and blue squares track the halves above and below the minor axis, which correspond to the receding and approaching halves respectively for these three galaxies (see Figure~\ref{fig:filteredvelmaps}).} The shading in the bottom panel indicates the approximate extent of a bar aligned along the major axis of UGC~9334, and thus where \Vrbar and \Vtbar in Equation~\ref{eq:radflowequation} are degenerate.
	\label{fig:triplegalradflows}
\end{figure}

One goal of the analysis presented here is to indirectly detect accretion by constraining radial flows, and this is where we begin. The poor data quality of UGC~8475 evident in Figure~\ref{fig:filteredvelmaps} prohibits detailed kinematic modelling, particularly in the inner disc where velocity uncertainties are high. Thus, we apply the radial flow model in Equation~\ref{eq:radflowequation} to the velocity and velocity-uncertainty maps of UGC~7899, UGC~9037, and UGC~9334, with the truncation radius set to the model extent in order to search for radial flows across the entire disc of each galaxy. The resulting \Vrbar from models of the entire map (black line), as well as of subsets of the data defined relative to the kinematic major and minor axes (coloured lines) of each galaxy are shown in Figure~\ref{fig:triplegalradflows}. As discussed in Section~\ref{subsec:kinematicmodels}, a non-zero \Vrbar does not necessarily imply physical radial inflows; rather we interpret a non-zero \Vrbar as a signature of non-circular flows to be more fully characterized in subsequent sections. We have verified that the radial flow profiles are unchanged if the disc geometries in the kinematic models are fixed to photometric values from \cite{huang2014highmass}; we therefore find it unlikely that the non-circular flows discussed below result from errors in the disc geometry.

\begin{figure}
	\centering
	\includegraphics[width=0.45\textwidth]{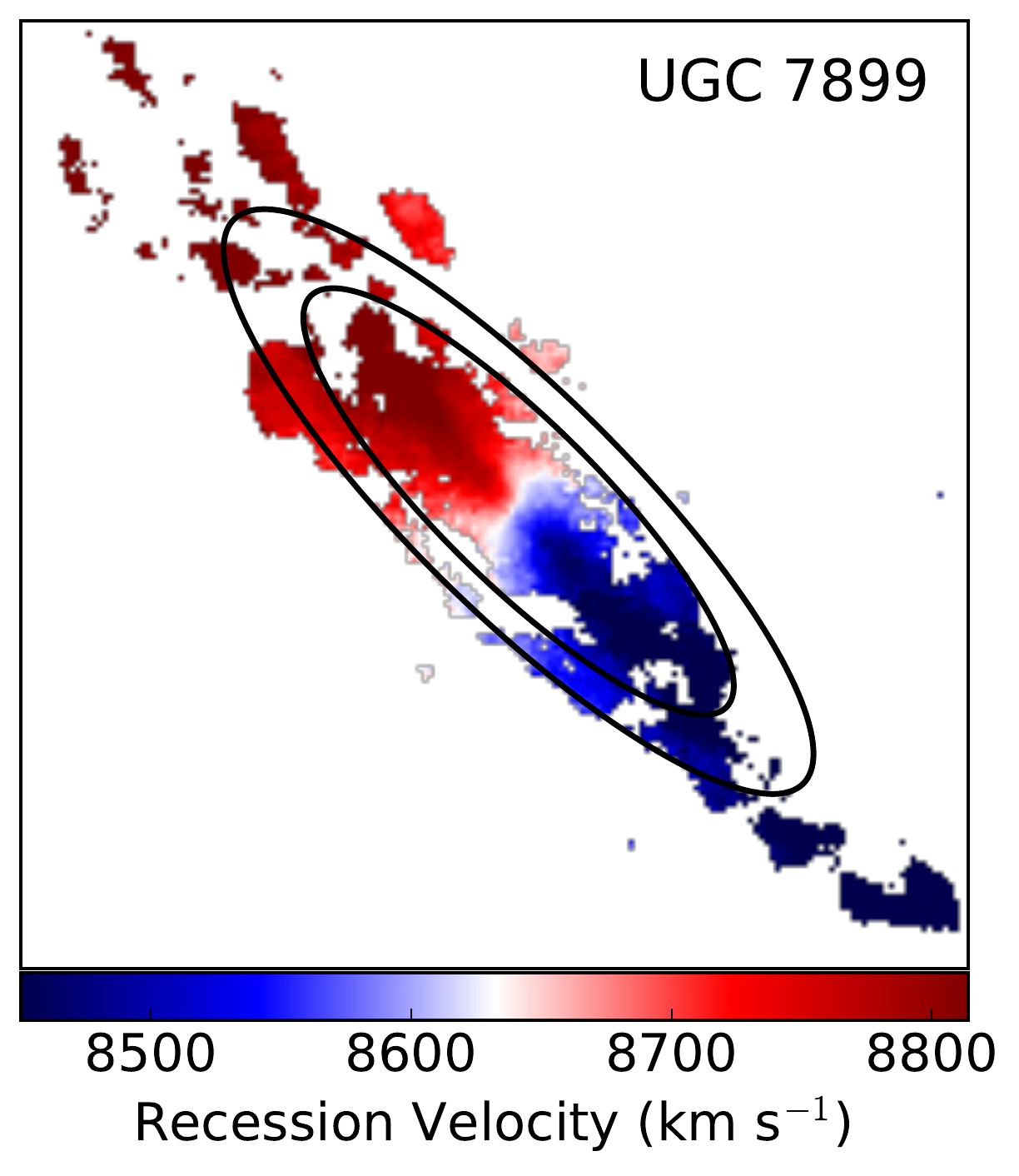}
	\caption[UGC 7899 Dip Location]{Annulus corresponding to $r = 10$~kpc (inner black ellipse) and $r = 15$~kpc (outer black ellipse) in the disc plane for the best-fitting kinematic model geometry in our preferred model of UGC~7899 (Table~\ref{table:fittedparameters}) superimposed on the final velocity map. There is a dearth of data North of the kinematic major axis in the map, which likely explains the discrepant \Vrbar relative to other radial flow models in Figure~\ref{fig:triplegalradflows} as well as the dip in the rotation curve for UGC~7899 in that region in Figure~\ref{fig:UGC7899RCfig}. See text for details.}
	\label{fig:UGC7899_ellipse_gap}
\end{figure}

The top panel of Figure~\ref{fig:triplegalradflows} shows that the radial flow models of UGC~7899 consistently find a non-zero \Vrbar for $r \leq 5$~kpc, suggesting that the model has uncovered coherent non-circular flows in the inner disc. Further out, there are hints of non-zero \Vrbar at $r \sim 10$~kpc and $15$~kpc~$\lesssim r \lesssim 20$~kpc, with model amplitudes as large as $50$~\kms in the latter region. However, these signatures are less coherent across models of different data subsets compared to \Vrbar in the inner disc. The most discrepant among subset models is that from the Northern half (blue line and squares), which we speculate arises from a dearth of data North of the major axis beyond $r=10$~kpc in the velocity map. This is illustrated in Figure~\ref{fig:UGC7899_ellipse_gap}, where sky-plane projections of the locations with $r=10$~kpc (inner black ellipse) and $r=15$~kpc (outer black ellipse) are superimposed on the velocity maps for UGC~7899. Models of the approaching segment (blue lines and squares in the top panel of Figure~\ref{fig:triplegalradflows}) also return $\Vrbar \sim 0$ for $r > 15$~kpc. We therefore conclude that there is a clear signature of non-circular flows for $r < 5$~kpc and marginal evidence for non-circular flows beyond this radius in UGC~7899. In Section~\ref{subsec:appofkinmodels}, we will model the non-circular flows in the inner disc as both radial flows and a bar-like flow and determine our preferred physical model.

The middle panel of Figure~\ref{fig:triplegalradflows} shows that the radial flow models of UGC~9037 consistently find a non-zero \Vrbar for $r \lesssim 3.5$~kpc, suggesting that these models, too, have uncovered coherent non-circular flows in the inner disc. On the other hand, all of the models return \Vrbar $\sim 0$ beyond the inner disc, albeit with some scatter between models of different data subsets. Adopting the uncertainties on the model applied to the velocity map as a whole (black line) as an approximate upper limit, our models imply that $\Vrbar < 15 \ \kms$ in the outer disc of UGC~9037. In Section~\ref{subsec:appofkinmodels}, we will model the non-circular flows in the inner disc as both radial flows and a bar-like flow and determine our preferred physical model.

The bottom panel of Figure~\ref{fig:triplegalradflows} shows that the best-fitting \Vrbar to the entire velocity map for UGC~9334 (black line) dips below 0 for $10\, \text{kpc} \lesssim r \lesssim 15$~kpc and above 0 for $20\,\text{kpc} \lesssim r \lesssim 25$~kpc. The latter flows are not consistently recovered in different subsets of the data. We therefore find no convincing signature of non-circular flows for $r \gtrsim 3$~kpc for UGC~9334, with the models suggesting that $\Vrbar < 20$~\kms on average across the disc. There is strong evidence that $\Vrbar < 0$ for $r \lesssim 3$~kpc in all of the models, which implies coherent non-circular flows in that region. We speculate that they stem from a bar aligned along the major axis of UGC~9334 \citep{willett2013galaxy} that is visible in the bottom panel of Figure~\ref{fig:quadimageSDSSDR14}. We highlight the approximate extent of the bar region in blue in the bottom panel of Figure~\ref{fig:triplegalradflows}, and note that the DiskFit's bisymmetric model (Equation~\ref{eq:m2perturbations}) is degenerate for this geometry (\citealt{randriamampandry2016exploring}; see Section~\ref{subsec:kinematicmodels}). In Section~\ref{subsec:appofkinmodels}, we will therefore apply rotation-only models to the velocity map of UGC~9334, with the caveat that they do not reliably trace the physical structure of this system in the bar region.

\subsection{Application of Kinematic Models}\label{subsec:appofkinmodels}

Informed by the radial flow search of the previous section, we present the rotation-only models to the velocity and velocity-uncertainty maps for all four galaxies, as well as inner bisymmetric and radial flow models for UGC~7899 and UGC~9037. Section~\ref{subsubsec:8475and9334} presents rotation-only models for UGC~8475 and UGC~9334, and Sections~\ref{subsubsec:7899} and \ref{subsubsec:9037} present applications of all three models to UGC~7899 and UGC~9037 respectively. The best-fitting geometries for all models are given in Table~\ref{table:fittedparameters}, and Figures~\ref{fig:UGC8475Vt}~$-$~\ref{fig:UGC9037RCfig} show the models, residuals, and kinematic components. In Section~\ref{subsec:noncircflows78999037} we discuss our preferred models for UGC~7899 and UGC~9037 among those applied to the data.

\begin{table*}
	\begin{tabular}{ccccccccc} \hline \hline
		Galaxy & Model & RA (J2000) [H:M:S] & DEC (J2000) [D:M:S] & \diskPA [\degreesign] & $i$ [\degreesign] & $r_t$ [kpc]  &  \phib [\degreesign] & \phibsky [\degreesign]  \\
		~ & (1) & (2) & (3) & (4) & (5) & (6) & (7) & (8) \\ \hline \hline
						 & Rotation Only & 12:43:45.50(6) & 10:46:13.74(6) & 45.4 $\pm$  0.6 & 74.3 $\pm$  0.8 & $\ldots$ & $\ldots$ & $\ldots$  \\
		UGC 7899 & Inner Bisymmetric & 12:43:45.50(4) & 10:46:13.74(4) & 45.4 $\pm$  0.3 & 74.4 $\pm$  0.6 & 4.9 & 33 $\pm$ 6 & 55, 23 $\pm$ 4 \\
						 & \textbf{Inner Radial Flow} & 12:43:45.50(6) & 10:46:13.74(5) & 45.2 $\pm$  0.5 & 74.0 $\pm$  0.7 & 4.9 & $\ldots$ & $\ldots$ \\ 
		\hline
		UGC 8475 & \textbf{Rotation Only} & 13:29:25.8(8) & 11:00:28.5(7) & 167 $\pm$ 1 & 62 $\pm$ 2 & $\ldots$ & $\ldots$ & $\ldots$ \\ 
		\hline
						 & Rotation Only & 14:08:29.3(2) & 07:03:28.4(2) & 345.0 $\pm$  0.8 & 63 $\pm$  2 & $\ldots$ & $\ldots$ & $\ldots$ \\
		UGC 9037 & \textbf{Inner Bisymmetric} & 14:08:29.3(1) & 07:03:28.2(1) & 344.6 $\pm$  0.6 & 61 $\pm$  1 & 3.3 & 24 $\pm$ 13 & 117, 177 $\pm$ 8 \\
						 & Inner Radial Flow & 14:08:29.3(2) & 07:03:28.2(2)  & 344.5 $\pm$  0.7 & 61 $\pm$  2 & 3.3 & $\ldots$ & $\ldots$ \\ 
		\hline
		UGC 9334 & \textbf{Rotation Only} & 14:31:01.1(1) & 05:58:42.3(1)  & 290.3 $\pm$  0.5 & 52.6 $\pm$  0.7 & $\ldots$ & $\ldots$ & $\ldots$
	\end{tabular}
	\caption{Fitted parameters for the fits performed on UGC~7899, UGC~8475, UGC~9037, and UGC~9334. Col 1: the model applied (preferred model in bold). Cols 2 and 3: right ascension and declination of the galaxy centre (uncertainty in the last digit shown in parentheses). Col 4: the position angle of the galaxy disc. Col 5: the inclination of the galaxy disc. Col 6: the truncation radius for a fit, only applicable to constrained non-circular flow fits. Col 7: the disc-plane bar angle, only relevant to fits for inner bisymmetric flows. Col 8: possible sky-plane major axis bar angles, calculated from \phib (Col 7) and $i$ (Col 4) using Equation~\ref{eq:sky_to_disk_conversion} (see text for details).}
	\label{table:fittedparameters}
\end{table*}

\subsubsection{UGC~8475 and UGC~9334}\label{subsubsec:8475and9334}

\begin{figure}
	\centering
	\includegraphics[width=0.45\textwidth]{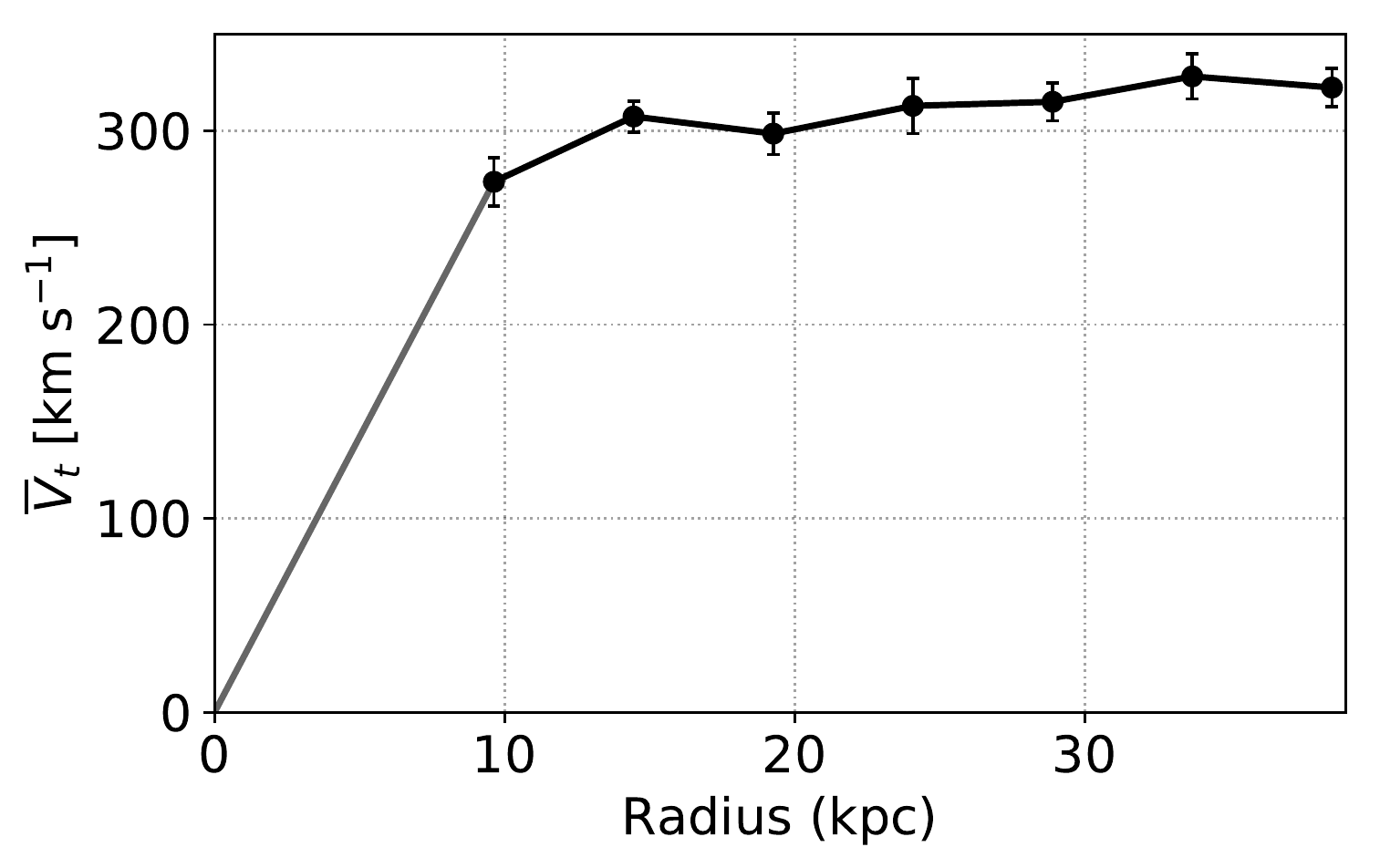}
	\caption[Rotation Curve for UGC~8475]{\Vtbar in rotation-only model for UGC~8475 applied to the velocity and velocity-uncertainty maps in Figure~\ref{fig:filteredvelmaps}.}
	\label{fig:UGC8475Vt}
	\vspace{-5.0pt}
\end{figure}

\begin{figure}
	\centering
	\includegraphics[width=0.45\textwidth]{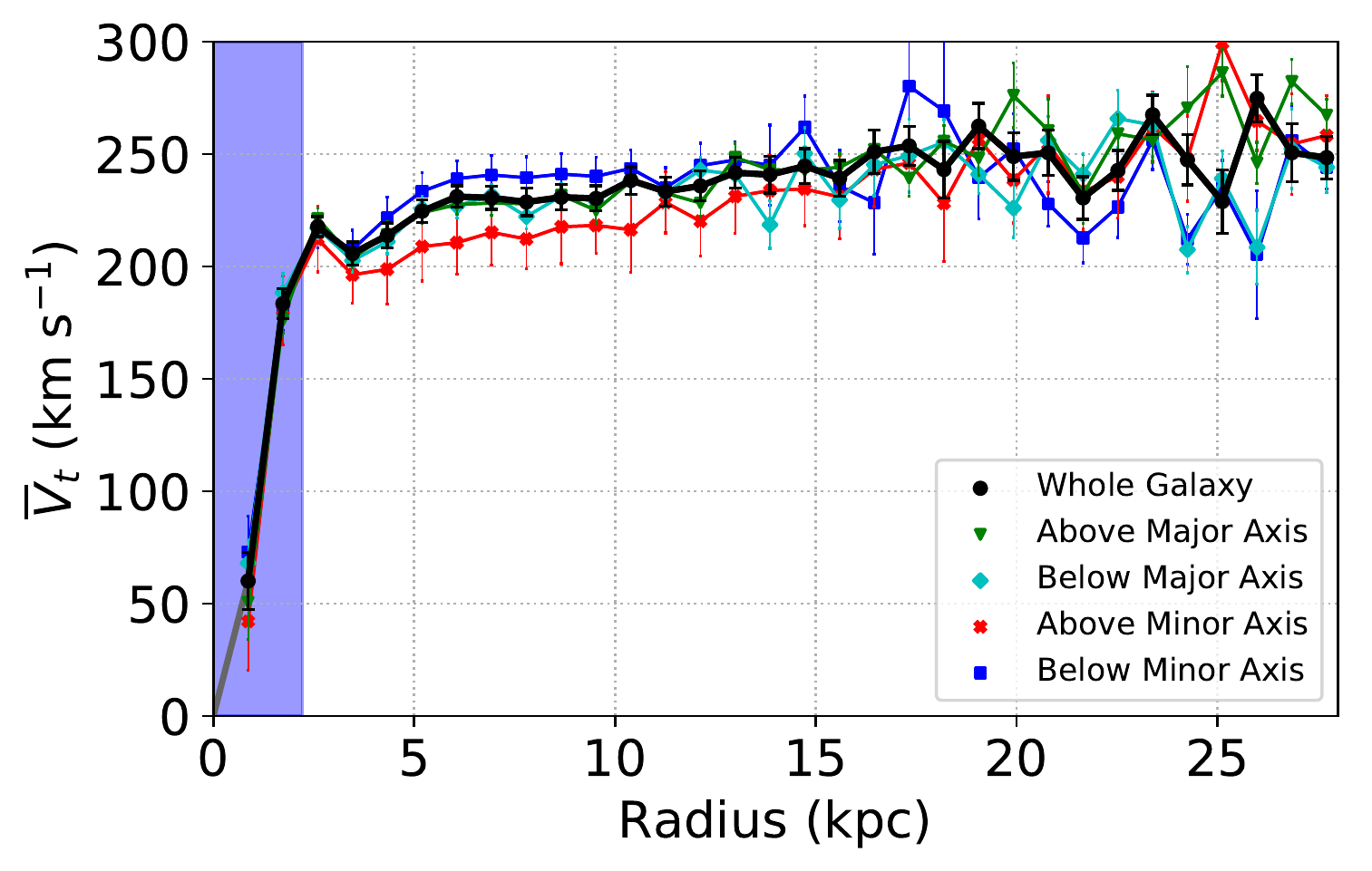}
	\caption[Rotation Curve for UGC~9334]{\Vtbar in rotation-only model for UGC~9334 applied to the velocity and velocity-uncertainty maps in Figure~\ref{fig:filteredvelmaps}. The black points show \Vtbar obtained by modelling the whole galaxy, whereas the coloured lines represent models run on different halves of the disc, relative to the kinematic major and minor axes of the whole-disc models, as described in the legend. For UGC~9334, the segment above the minor axis corresponds to the receding half of the disc, and the segment below the minor axis corresponds to the approaching half. The shaded region is the same as in Figure~\ref{fig:triplegalradflows}, and highlights the region where \Vtbar does not reliably trace rotation because of a bar aligned along the major axis (see Figure~\ref{fig:quadimageSDSSDR14}).}
	\label{fig:UGC9334Vt}
\end{figure}

As discussed in Section~\ref{subsec:ncflowsalldisk}, the sparsity of the UGC~8475 data and the bar along the UGC~9334 major axis limit our analysis of the velocity and velocity-uncertainty maps in Figure~\ref{fig:filteredvelmaps} for these objects to the application of rotation-only models. Their best-fitting disc geometries are given in Table~\ref{table:fittedparameters}, which are in good agreement with photometric measurements \citep{huang2014highmass}. The best-fitting \Vtbar for these models are given in Figure~\ref{fig:UGC8475Vt} for UGC~8475 and by the black line in Figure~\ref{fig:UGC9334Vt} for UGC~9334. To the best of our knowledge, Figures~\ref{fig:UGC8475Vt} and \ref{fig:UGC9334Vt} present the first rotation curve measurements for these galaxies in any tracer. Both curves are flat or mildly rising out to the last measured point, and their amplitudes are consistent with estimates from single-dish \Hone profile widths and photometric disc inclinations \citep{huang2014highmass}. We note that the \Halpha rotation curves of massive galaxies are generally flat (e.g. \citealt{catinella2006template}), and therefore that the mild rotation curve rise in both galaxies could stem from a missed warp in our flat disc models. However, the warp would need to begin well within the edge of the optical disc in both galaxies, which we consider unlikely.

The coloured lines in Figure~\ref{fig:UGC9334Vt} show \Vtbar obtained from rotation-only models applied to halves of the UGC~9334 velocity map in Figure~\ref{fig:filteredvelmaps}, defined from the kinematic major and minor axes of the whole-disc model given in Table~\ref{table:fittedparameters}. Models for the (statistically independent) Northern and Southern halves agree well within uncertainty across the disc, and all of the curves are generally consistent for $r > 15$~kpc. However, for $3 \ \text{kpc} < r < 10$~kpc, \Vtbar for the receding (above minor axis) half of UGC~9334 consistently exceeds that for the approaching (below minor axis) half by $\sim30 \kms$, suggesting a degree of kinematic lopsidedness. Relatively isolated galaxies do sometimes exhibit morphological and kinematic asymmetries (e.g. \citealt{greisen2009aperture}; \citealt{reynolds2020h}), and the origin of this feature in UGC~9334 deserves further study that we reserve for future work.

\subsubsection{UGC 7899}\label{subsubsec:7899}

\begin{figure}
	\centering
	\includegraphics[width=0.45\textwidth]{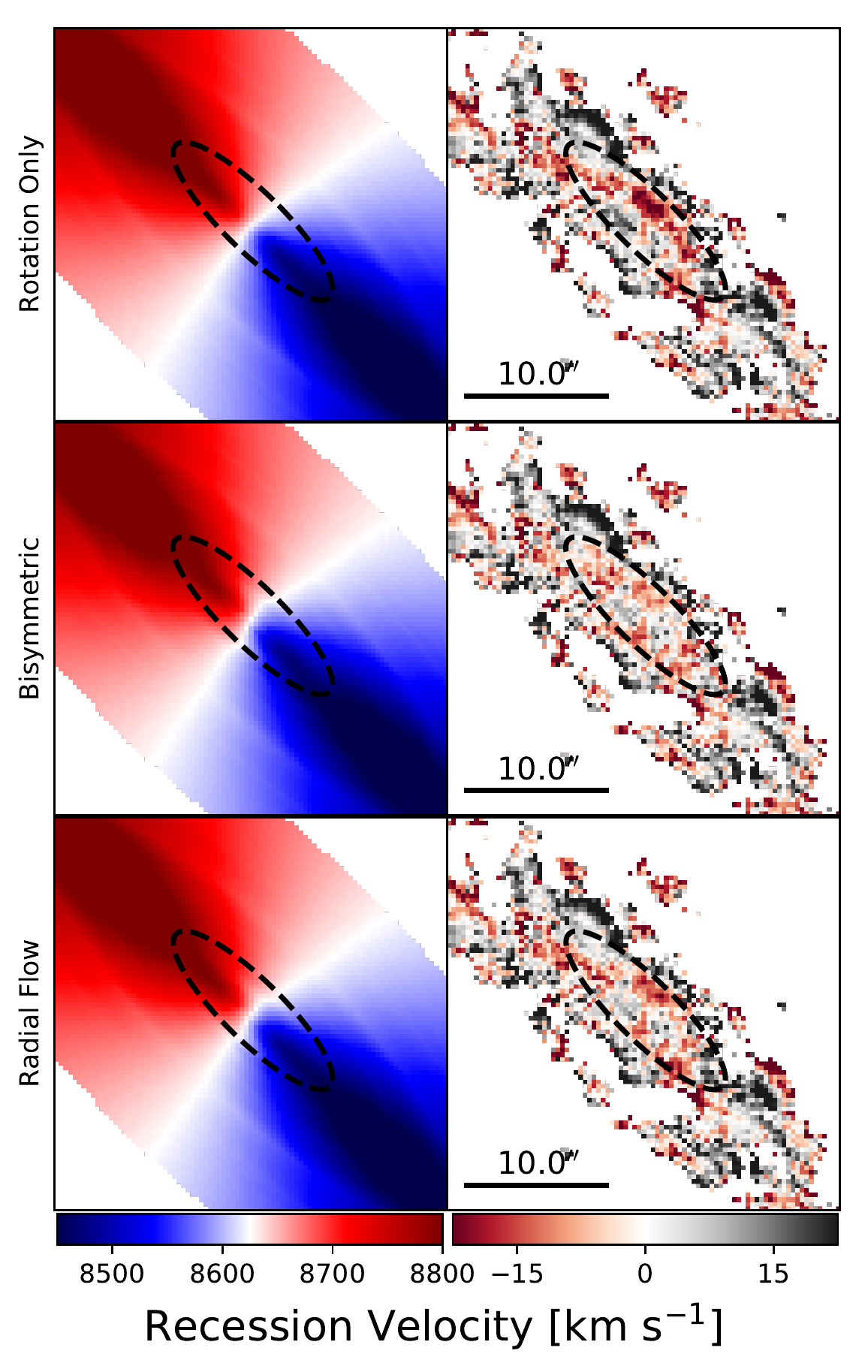}
	\caption[Model and Residuals for UGC 7899]{Best-fitting model (left) and residuals (right) for the inner disc for the rotation-only (top), inner bisymmetric (middle), and radial flow (bottom) models applied to the velocity and velocity-uncertainty maps for UGC~7899 in Figure~\ref{fig:filteredvelmaps}. A black, dashed ellipse extends to the truncation radius $r_t$ (only physically relevant for the bisymmetric and radial flow models but plotted on the rotation-only model for consistency), within which the discrepancy between the rotation-only and non-circular flow models is most severe. The black bar in the residual panels shows the angular scale, and the colour bars show the velocity scale in each column.}
	\label{fig:UGC7899MRfig}
\end{figure}

\begin{figure}
	\centering
	\includegraphics[width=0.45\textwidth]{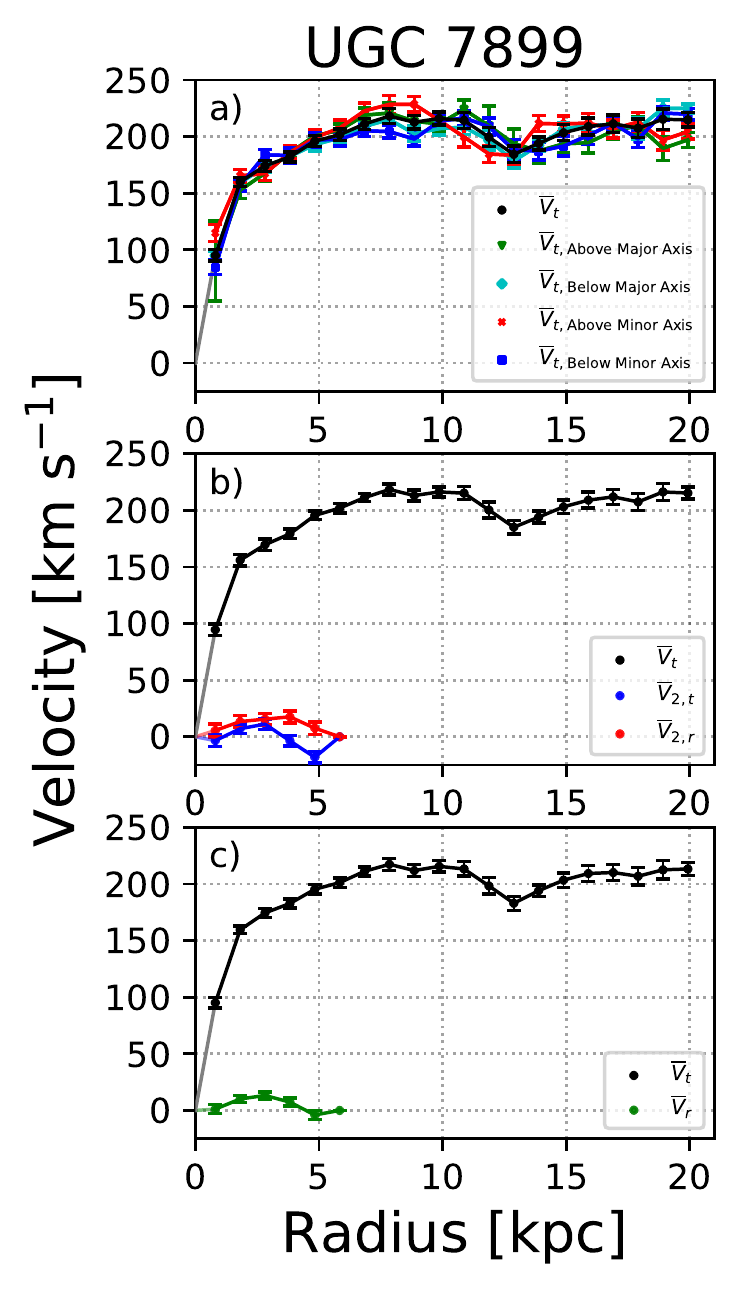}
	\caption[Model Rotation Curves for UGC 7899]{Best-fitting kinematic components from a) rotation-only, b) bisymmetric, and c) radial flow models for UGC~7899 shown in Figure~\ref{fig:UGC7899MRfig}. In all panels, the black points correspond to \Vtbar obtained by applying the model to the entire velocity map. In a), the coloured lines represent \Vtbar obtained from models run on different halves of the disc, relative to the kinematic major and minor axes of the whole-disc models, as described in the legend, with marker symbols consistent with Figure~\ref{fig:triplegalradflows}. In b), the blue and red lines correspond to the tangential and radial components of the bisymmetric flow, \Vtbarbar and \Vrbarbar respectively. In c), the green line corresponds to the inner-disc radial flow, \Vrbar. The $\sim$35 \kms amplitude dip in \Vtbar at $r \sim13$~kpc in all models is likely due to a lack of data within that region of the disc (see Figure~\ref{fig:UGC7899_ellipse_gap} and text for details).}
	\label{fig:UGC7899RCfig}
\end{figure}

Informed by the whole-disc radial flow models of Section~\ref{subsec:ncflowsalldisk}, we apply rotation-only, inner bisymmetric and inner radial flow models to the velocity and velocity-uncertainty maps for UGC~7899 in Figure~\ref{fig:filteredvelmaps}. In all models, we set the truncation radius for the non-circular flows as $r_t=4.9\,$kpc, which corresponds to the approximate extent of the coherent feature in the top panel of Figure~\ref{fig:triplegalradflows} in integer pixel units; the models do not depend strongly on this choice, with values in the range $4.5 \ \text{kpc} < r_t < 5.5$~kpc returning similar results. Figure~\ref{fig:UGC7899MRfig} presents the best-fitting models and residuals, Figure~\ref{fig:UGC7899RCfig} shows the kinematic components from the models, and Table~\ref{table:fittedparameters} reports the best-fitting disc geometries. 

Figure~\ref{fig:UGC7899MRfig} shows that by eye, the data-model residuals in the inner disc of UGC~7899 are both larger and more coherent for the best-fitting rotation-only model than for either the best-fitting inner bisymmetric or inner radial flow models. The latter two models are nearly indistinguishable in the sky plane, and the slightly larger residuals for the radial flow model relative to the bisymmetric model are expected from its smaller number of fitted parameters. 

The kinematic components in Figure~\ref{fig:UGC7899RCfig} and the disc geometries in Table~\ref{table:fittedparameters} are well-constrained in all three models. Consistent with the findings of \cite{holmes2015incidence} for the CALIFA sample, the disc geometries and axisymmetric tangential components \Vtbar between different models are identical to within uncertainty: all models return the same disc geometry and rotation curves. The disc geometries are consistent with previous photometric estimates from \textit{R}-band imaging (\citealt{huang2014highmass}, see also Section~\ref{subsec:noncircflows78999037}) and kinematic measurements from models applied to \Hone maps \citep{hallenbeck2016highmass}.

We find a $\sim35 \kms$ dip in \Vtbar at $r \sim 13\,$kpc  in all three models, as well as in rotation-only models run on different subsets of the data relative to the kinematic major and minor axes defined by the disc geometry in Table~\ref{table:fittedparameters}. We attribute this dip to a combination of UGC~7899 velocity map properties from $10$~kpc~$< r < 15$~kpc in the disc plane, a region highlighted by the black sky-plane ellipses in Figure~\ref{fig:UGC7899_ellipse_gap}. First, there is a clear dearth of emission in the Northern (above major axis) half of the disc: this likely underpins the change in \Vtbar for the Northern half of the galaxy (green line and upside-down triangles in upper panel of Figure~\ref{fig:UGC7899RCfig}) and the discrepancy in \Vrbar in the whole-disc radial flow models (green line and upside-down triangles in upper panel of Figure~\ref{fig:triplegalradflows}) for that subset. The gas signature picks up beyond $r > 15$~kpc along the major axis in particular, resulting in a rotation curve which recovers to its previous plateau. Second, Figure~\ref{fig:UGC7899_ellipse_gap} shows some evidence for a disconnect between the gas in the Southern half of the disc in that region and the emission at smaller $r$ that may connect to the tail-like structure to the South-West of the galaxy. Ultimately, the dip in \Vtbar at $r \sim 13\,$kpc most likely stems from a complexity in the kinematics that is missing in our models rather than a change in the underlying mass distribution of UGC~7899 in the \Halpha tracer. This may correspond to a warp in the disc, although it would lie well within the outer \Hone disc warp suggested by \cite{hallenbeck2016highmass} as well as the edge of the stellar disc \citep{huang2014highmass}.

Figure~\ref{fig:UGC7899RCfig} also shows that the inner non-circular components \Vtbarbar and \Vrbarbar in the bisymmetric model and \Vrbar in the radial flow model are both well-constrained, with similar extents, amplitudes and uncertainties. Combined with the similarities between the models, residuals and \Vtbar from Figures~\ref{fig:UGC7899MRfig} and \ref{fig:UGC7899RCfig}, we therefore find that the inner bisymmetric model and the inner radial flow model describe the kinematics of UGC~7899 equally well. We discuss our preferred model between the two in Section~\ref{subsec:noncircflows78999037}, and its physical implications in Section~\ref{sec:discussion}.

\subsubsection{UGC 9037}\label{subsubsec:9037}

\begin{figure}
	\centering
	\includegraphics[width=0.45\textwidth]{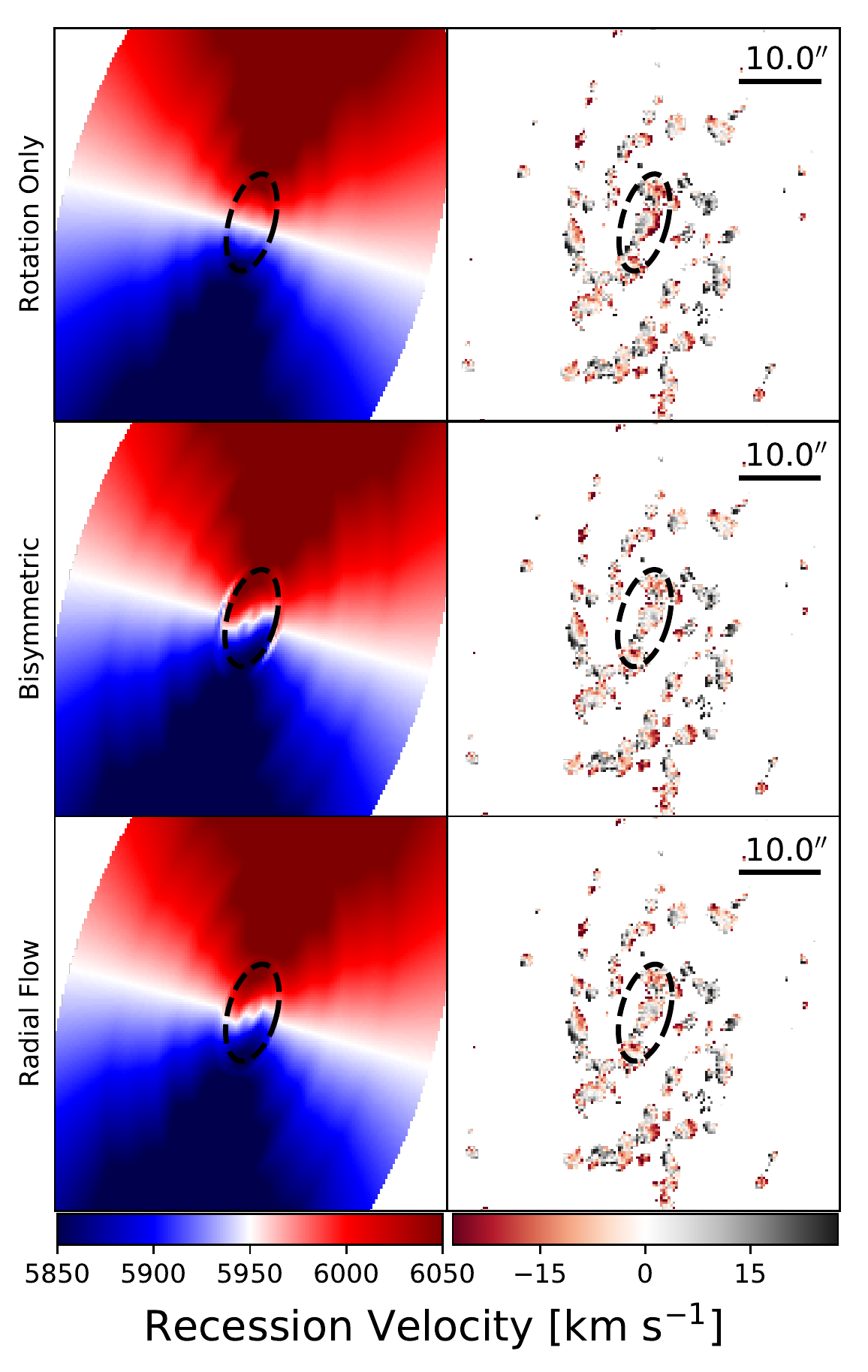}
	\caption[Model and Residuals for UGC 9037]{Same as in Figure~\ref{fig:UGC7899MRfig} but for UGC~9037.}
	\label{fig:UGC9037MRfig}
\end{figure}

\begin{figure}
	\centering
	\includegraphics[width=0.45\textwidth]{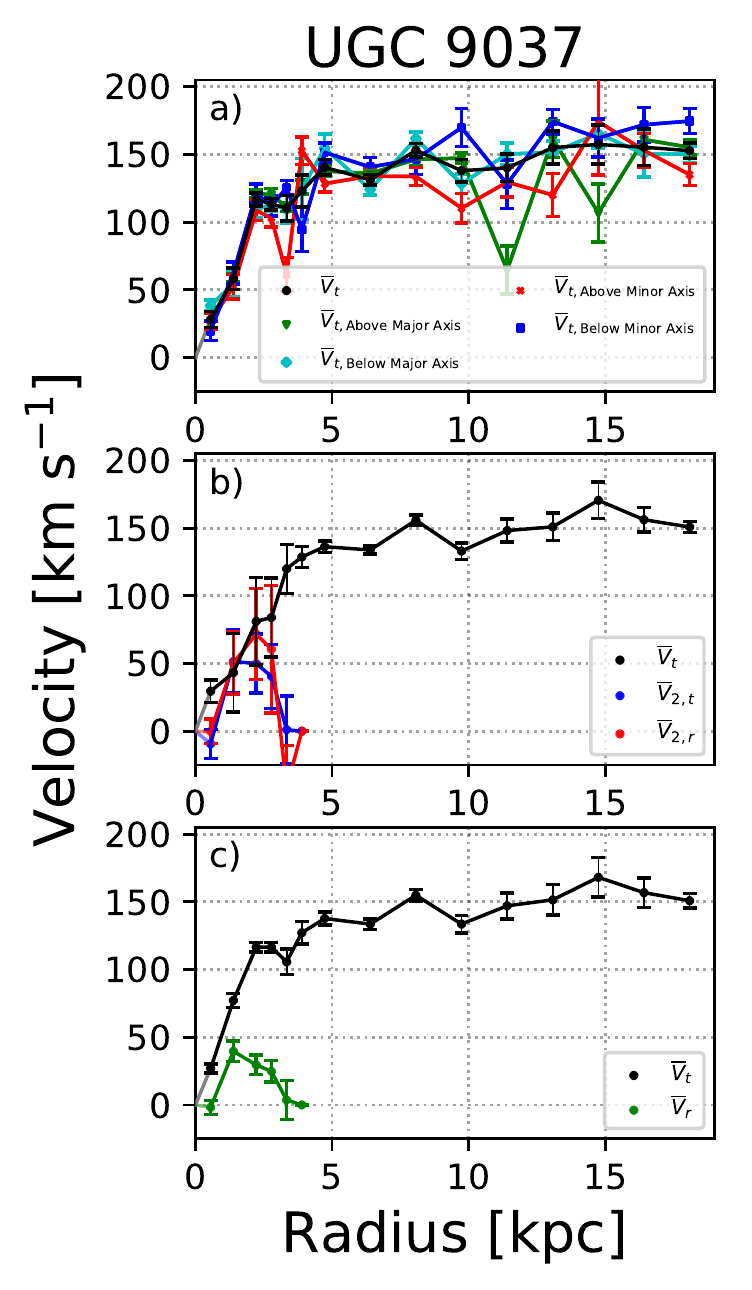}
	\caption[Model Rotation Curves for UGC 9037]{Same as in Figure~\ref{fig:UGC7899RCfig}, but for UGC 9037.}
	\label{fig:UGC9037RCfig}
\end{figure}

Informed by the whole-disc radial flow models of Section~\ref{subsec:ncflowsalldisk}, we apply rotation-only, inner bisymmetric and inner radial flow models to the velocity and velocity-uncertainty maps for UGC~9037 in Figure~\ref{fig:filteredvelmaps}. In all models, we set the truncation radius for the non-circular flows as $r_t=3.3\,$kpc, which corresponds to the approximate extent of the coherent feature in the top panel of Figure~\ref{fig:triplegalradflows} in integer pixel units; the models do not depend strongly on this choice, with values in the range $3 < r_t < 4$~kpc returning similar results. Figure~\ref{fig:UGC9037MRfig} presents the best-fitting models and residuals, Figure~\ref{fig:UGC9037RCfig} shows the kinematic components from the models, and Table~\ref{table:fittedparameters} reports the best-fitting disc geometries. 

The kinematic components in Figure~\ref{fig:UGC9037RCfig} and the disc geometries in Table~\ref{table:fittedparameters} are well-constrained in all three models. The disc geometries are consistent with previous photometric estimates from \textit{R}-band imaging (\citealt{huang2014highmass}, see also Section~\ref{subsec:noncircflows78999037}) and kinematic measurements from models applied to \Hone and CO maps \citep{hallenbeck2014highmass}. The coloured lines in Figure~\ref{fig:UGC9037RCfig}a) show \Vtbar obtained by fitting halves of the data relative to the kinematic major axis of the whole disc model as described in the legend. The models for these data subsets are generally consistent with one another, particular for $r < 10$~kpc. 

Figure~\ref{fig:UGC9037MRfig} shows that by eye, the data-model residuals in the inner disc of UGC~9037 are both larger and more coherent for the best-fitting rotation-only model than for either the best-fitting inner bisymmetric or inner radial flow models. The larger uncertainties in the inner bisymmetric flows (as compared to the inner radial flow model) are characteristic of bootstrap resampling with a model that has more free parameters (e.g. \citealt{holmes2015incidence}). As is the case for UGC~7899, the best-fitting inner bisymmetric and inner radial flow models are nearly indistinguishable in the sky plane. The similarities between the models, residuals and \Vtbar from Figures~\ref{fig:UGC9037MRfig} and \ref{fig:UGC9037RCfig} respectively, preclude drawing conclusions from kinematic modelling alone. We discuss our preferred model for UGC~9037 in Section~\ref{subsec:noncircflows78999037}, and its physical implications in Section~\ref{sec:discussion}.

\subsection{Preferred models for UGC~7899 and UGC~9037}\label{subsec:noncircflows78999037}

Our analysis of the velocity and velocity-uncertainty maps for UGC~7899 and UGC~9037 in Section~\ref{subsubsec:7899} and \ref{subsubsec:9037} proved inconclusive with regard to determining the ideal kinematic model for either system. Thus, we combine our kinematic analysis with photometric modelling to determine our preferred physical model. We use the photometric branch of DiskFit to model the \textit{R}-band imaging for UGC~7899 and UGC~9037 presented in \cite{huang2014highmass}. 

The photometric branch of DiskFit, which uses the same optimization algorithm as for the kinematic branch, can decompose images into disc, bar and bulge components \citep{kuzio2012searching, sellwood2015diskfit, holmes2015incidence, lewis2018non}. The flat disc and fixed-angle bar are described by non-parametric surface brightness profiles and the bulge has a S\'ersic functional form. A disc-bar-bulge model returns the best-fitting photometric disc position angle \diskPAphot, inclination $i_{\text{phot}}$, and the sky-plane bar position angle \phibskyphot.

We apply disc-bar-bulge models to the \textit{R}-band imaging for UGC~7899 and UGC~9037, using 1000 bootstrap realizations to estimate uncertainties. The best-fitting parameters from these runs are reported in Table~\ref{table:photometricparameters}, and can be compared with their kinematically derived counterparts in Table~\ref{table:fittedparameters}. We determine our preferred kinematic model by requiring that the implied structure is broadly consistent with our best-fitting photometric decompositions.

\begin{table}	
	\begin{tabular}{rccc} \hline \hline
		 & \diskPAphot [\degreesign] & \iphot [\degreesign] & \phibskyphot [\degreesign] \\ 
		 Galaxy & (1) & (2) & (3) \\ \hline
		UGC~7899   &  42.6 $\pm$ 0.7 & 76.6 $\pm$ 0.7 & 36 $\pm$ 2 \\ 
		UGC~9037   &  330 $\pm$ 6 & 48 $\pm$ 5 & 112 $\pm$ 3 \\
		\hline     
	\end{tabular}
	\caption[Best-fitting photometric parameters for UGC~7899 and UGC~9037]{Best-fitting parameters derived from the photometric disc-bar-bulge model applied to \textit{R}-band images of UGC~7899 and UGC~9037. Col 1: the disc position angle. Col 2: the disc inclination. Col 3: the sky-plane bar angle.}
	\label{table:photometricparameters}
	\vspace{-5.5pt}
\end{table}

Comparing \phibsky and \phibskyphot for UGC~7899 in Tables~\ref{table:fittedparameters} and \ref{table:photometricparameters} respectively, we find that the sky-plane position angle of the bar implied by our photometric models is inconsistent with either possible \phibsky calculated from the inner bisymmetric kinematic model of this galaxy (Equation~\ref{eq:sky_to_disk_conversion}). In the case of UGC~9037 on the other hand, the kinematic sky-plane bar position angle $\phibsky = (117 \pm 8)\degreesign$ (Table~\ref{table:fittedparameters}), is consistent to within uncertainty with its photometric sky-plane bar position angle, $\phibskyphot = (112 \pm 3)\degreesign$ (Table~\ref{table:photometricparameters} and Figure~\ref{fig:quadimageSDSSDR14}), implying that a bar at this position angle explains both the \textit{R}-band photometry and inner \Halpha kinematics for this galaxy.

We note that \diskPAphot and \iphot for UGC~9037 differ from their kinematic models at approximately twice the model standard deviation implied by the uncertainties on the fit. As a check, we re-ran the disc-bar-bulge photometric models for UGC~9037 with the disc geometry fixed to the kinematic values and recover a \phibskyphot that is consistent with the one reported in Table~\ref{table:photometricparameters}. This is not unexpected since the photometric disc geometry in a flat-disc model is typically driven by the outer disc morphology, while the photometric bar geometry is driven by that of the inner disc.

The lack of consistency between the disc geometry and bar angle for the inner bisymmetric flow model and for the disc-bulge-bar decomposition for UGC~7899 implies that we cannot find a bar-like model that consistently reproduces its kinematics and morphology. We therefore favour inner radial flows to explain the non-circular motions in UGC~7899. Conversely, our kinematic and photometric models of UGC~9037 can be consistently modelled with an inner bar, and we favour this interpretation for its underlying structure.

\section{Discussion}\label{sec:discussion}

In this Section, we discuss the implications of the kinematic models presented in Section 3 for the physical structure of the HIghMass galaxies and the late accretion hypothesis. We place our preferred models for UGC~7899 and UGC~9037 in the context of other constraints on their structure in Section~\ref{subsec:compwithlit}. In Section~\ref{subsec:radialflowsandlateaccretion}, we consider our constraints on disc-wide radial flows for UGC~7899, UGC~9037, and UGC~9334 in the context of hot and cold gas accretion models.

\subsection{The Structure of UGC~7899 and UGC~9037}\label{subsec:compwithlit}

While both our inner radial flow models and inner bisymmetric models for UGC~7899 and UGC~9037 adequately describe the measured \Halpha kinematics for these systems (Sections~\ref{subsubsec:7899} and \ref{subsubsec:9037} respectively), a comparison between the implied inner-disc structure with photometric disc-bar-bulge decompositions suggest inner radial flows in UGC~7899 and inner bisymmetric flows in UGC~9037. Here, we discuss those interpretations in the context of other constraints on those systems as well as in the broader gas-rich galaxy population.

Our preferred model for UGC~7899 is that of a disc galaxy with inner radial flows and marginal evidence for non-circular flows beyond the inner disc. We therefore do not corroborate consistent $\Vrbar \sim 20 \ \kms$ radial flows for $10 \lesssim r \lesssim 20$~kpc suggested by some of the \Hone modelling presented by \cite{hallenbeck2016highmass}. However, our results are still consistent with the overall picture of this galaxy's morphology, structure, and star formation from that study.

In both \Hone and \Halpha, UGC~7899's \Halpha distribution is visibly asymmetric (see Figures~\ref{fig:quadimageSDSSDR14} and \ref{fig:filteredvelmaps}). Emission at the northern, receding edge is more extended than the southern, approaching lobe, which narrows (see Figure~\ref{fig:verticalreganderrUGC7899}). \cite{hallenbeck2016highmass} also show that the position angle of the \Hone disc changes beyond $r \sim 20$~kpc. These anisotropies may indicate an \Hone warp, and the disconnect between the gas in the Southern half of the disc and the emission at smaller radii for $10 \ \text{kpc} < r < 15 \ \text{kpc}$ (Figure~\ref{fig:UGC7899_ellipse_gap}) may be related. We return to the warp and outer disc kinematics in UGC~7899 in the context of late accretion in Section~\ref{subsec:radialflowsandlateaccretion}.

Under the assumption of trailing spiral arms (\citealt{lindblad1963possibility,contopoulos1971preference}), we can determine the direction of fitted radial flows in UGC~7899 as per Equation~\ref{eq:radflowequation}. The spiral arm structure of UGC~7899 suggests that its motion is clockwise in the plane of the sky. This implies that the Eastern semi-minor axis ($\theta = 90\degreesign$) is the nearest point of the disc, and therefore that $\Vrbar > 0$ detected represents inflows.

\cite{hallenbeck2016highmass} found that the surface densities of \Hone and \Hmol gas reach values of $30 \ \Msolarperpcsq$ and $10 \ \Msolarperpcsq$ respectively in UGC~7899, implying that this gas is marginally unstable within $r < 30$~kpc, and strongly unstable for $r < 10$~kpc. Since UGC~7899 is observed to most actively form stars in its inner regions, this is consistent with inner radial inflow. UGC~7899 has a typical SFR for its \Hmol mass, and its star formation efficiency drops as a function of disc radius, suggesting that the bottleneck for star formation is the atomic to molecular (\Hone-to-\Hmol) conversion pathway \citep{hallenbeck2016highmass}. This in turn suggests imminent star formation in the central regions.

If the radial flows we detect are fuelling this inner-disc star formation, then their peak amplitude $\Vr \sim 10 \ \kms$ at $r \sim 2$~kpc for UGC~7899 in Figure~\ref{fig:UGC7899RCfig}c allows for an estimate of the gaseous mass influx $\mu$ for a thin axisymmetric disc via $\mu = 2 \pi r \Sigma_g \Vr$. If the flows detected in our \Halpha maps are present in the molecular and atomic phases, then adopting a characteristic $\Sigma_g \sim 40 \ \Msolarperpcsq$ \citep{hallenbeck2016highmass} implies a mass inflow $\mu \sim 5 \ \Msolarperyear$ or $\Sigma_{\mu} \sim 0.4 \ \Msolarperyear \ \text{kpc}^{-2}$ within that radius. If this gas will be turned into stars to avoid the continuity problem implied by radial flows \citep{wong2004search,spekkens2007modeling}, then the corresponding star formation rate and central surface density exceeds estimates from \Halpha imaging for UGC~7899 and other HIghMass galaxies \citep{huang2014highmass}. This further supports the conclusion that our preferred model for the inner non-circular flows implies an imminent phase of significant central star formation.
	
Our results for UGC~9037 favour inner-disc non-circular flows that can be explained by the bar that is visible in images of this galaxy (Figure~\ref{fig:quadimageSDSSDR14}). The extent and amplitude of the inner bisymmetric flow are broadly commensurate with that found in \Halpha in CALIFA galaxies \citep{holmes2015incidence}. We do not corroborate the widespread non-circular flows farther out identified in some \Hone kinematic models in \cite{hallenbeck2014highmass}, instead placing an upper limit $\Vrbar \lesssim 15 \ \kms$ on their amplitude for $r \gtrsim 5$~kpc (Section~\ref{subsec:ncflowsalldisk}). We discuss the implications of this limit in the context of late accretion in Section~\ref{subsec:radialflowsandlateaccretion}.

While some \Hone models for UGC~9037 in \cite{hallenbeck2014highmass} also suggested the presence of non-circular flows in the inner disc, they are distinct from the ones that we detect in \Halpha. Importantly, the orientation of the central minor-axis twist in the \Hone velocity maps (Figure 2 in \citealt{hallenbeck2014highmass}) is opposite the one evident in the non-circular flow models of the \Halpha velocity maps presented here (Figure~\ref{fig:UGC9037MRfig}). When these signatures are both modelled as radial flows, the \Hone feature represents inflow while the \Halpha feature represents outflow if the spiral arms in UGC~9037 are trailing.
		
While we prefer a bisymmetric flow to explain the inner non-circular flows that we detect in UGC~9037 because of their consistency with photometric models, we also note that the outflows implied by our inner disc radial flow models are challenging to reconcile with constraints from other tracers. Gas outflows can plausibly result from intense star formation \citep{nath2009starburst}, but we have no evidence for such an event in UGC~9037. 

The specific star formation rate and central \Hone surface density of UGC~9037 are both greater than the typical value for a galaxy with the same stellar mass \citep{huang2014highmass,hallenbeck2014highmass}, and a variety of axisymmetric instability indicators (see Section~\ref{sec:intro}) suggest that the \Hone phase is marginally unstable in the inner region as well. These past observations indicate that UGC~9037 may be on the verge of a significant central star formation phase. Our preferred kinematic model for UGC~9037 - that of an axisymmetric outer disc with an inner bar - may be consistent with this basic picture, since bars can efficiently shuttle gas into the galaxy interior \citep{oh2011bar}. However, since bars are long-lived this scenario implies that we are observing the system at a special time, and it is therefore more likely that the bar and central star formation potential are disconnected.

While galaxies with strong bars tend to be less gas rich than those without \citep{masters2012galaxy}, our kinematic model for UGC~9037 as well as the bar in UGC~9334 that precludes a search for inner non-circular flows (see Figure~\ref{fig:quadimageSDSSDR14} and Section~\ref{subsec:ncflowsalldisk}) suggest that weak bars may be as common in massive gas-rich galaxies just as they are in the broader galaxy population (e.g. \citealt{erwin2018dependence}). Notwithstanding the small sample size considered here, this correspondence may suggest that massive, gas-rich galaxies evolve secularly in a similar manner to other galaxies.

\subsection{Radial Flows and Late Accretion} \label{subsec:radialflowsandlateaccretion}

We now turn to the implications of our constraints on non-circular flows in UGC~7899, UGC~9037 and UGC~9334 for the late accretion hypothesis. As outlined in Section~\ref{sec:intro}, cosmological simulations suggest hot-mode accretion as the dominant mechanism in the HIghMass galaxies given their relatively high stellar masses. On the other hand, the high gas-richness of the HIghMass galaxies raises the possibility that recent cold-mode accretion played a role in building their \Hone reservoirs. We discuss both possibilities below.

To examine the hot-mode accretion scenario, we follow the analytic framework described in \cite{pezzulli2016accretion}, and estimate the accretion rate surface density in the outer disc as

\begin{equation} \label{eq:accratesigma}
\dot{\Sigma}_{\text{acc}} = \frac{\Sigma_g \Vr}{\alpha r}   \ \text{,}
\end{equation}

\noindent where $\alpha$ is a unitless parameter measuring the local angular momentum deficit of the infalling material relative to the corotating disc flow.  For $\alpha \gtrsim 0.2$ the corresponding disc radial flow amplitude grows with increasing $r$, and we therefore use the results of the disc-wide radial flow searches in Section~\ref{subsec:ncflowsalldisk} to constrain the accretion due to radial inflows in the outer discs of UGC~9037 and UGC~9334 which may have gone undetected at the resolution and sensitivity of our observations.

Neither galaxy shows evidence for radial flows in their outer discs (see Figure~\ref{fig:triplegalradflows}). Adopting $\Vrbar < 20 \ \kms$ at $r \sim 15$~kpc appropriate for UGC~9334 as a conservative upper limit on the radial inflow amplitude that we can detect, a total disc gas surface density of $\Sigma_g(r \ = \ 15 \ \text{kpc}) \sim 10 \Msolarperpcsq$ (e.g. \citealt{hallenbeck2014highmass}), and an angular momentum mismatch $\alpha = 0.3$ on the higher end of that considered by \cite{pezzulli2016accretion}, Equation~\ref{eq:accratesigma} implies an upper limit on the accretion rate surface density of $\dot{\Sigma}_{\text{acc}} < 45 \ \Msolar \ \text{pc}^{-2} \ \text{Gyr}^{-1}$ in UGC~9037 and UGC~9334. This is substantially higher than the current star formation rate densities of HIghMass galaxies \cite{huang2014highmass}, as well as that of the Milky Way. Furthermore, given the inverse relationship betwen $\dot{\Sigma}_{\text{acc}}$ and $\alpha$, our outer disc radial flow constraints allow for substantially higher $\dot{\Sigma}_{\text{acc}}$ values than estimated here for smaller angular momentum mismatches. Thus, while we do not find evidence for outer disc radial flows in UGC~9037 and UGC~9334, we have found weak constraints on hot-mode accretion via these implied upper limits on outer disc radial flows. In particular, the predicted accretion rates for the Milky Way are much smaller than we can detect in the velocity maps, and as such they are consistent with our data.

In Section~\ref{subsec:noncircflows78999037}, we advocated radial flows as the preferred model for the inner disc ($r \ \lesssim \ 5$~kpc) kinematics of UGC~7899, and in Section~\ref{subsec:compwithlit}, we argued that inflows of the measured amplitude are consistent with other observations which suggest imminent central star formation in this galaxy. We now discuss possible origins of these flows. The inconsistency between the radial dependence of the flow constraints in UGC~7899 and that expected from the hot-mode accretion models of \cite{pezzulli2016accretion} suggests that, if late accretion is at work, then cold-mode accretion is the more likely explanation. This latter scenario would require a significant mass flow of cold gas to reach the inner disc of UGC~7899 (Section~\ref{subsec:compwithlit}, \citealt{hallenbeck2016highmass}).

Unlike UGC~9037 and UGC~9334, UGC~7899 does have hints of outer disc radial flows that broadly increase with radius (see ~\ref{fig:triplegalradflows}), as well as evidence for a disconnect between the detected \Halpha emission at $10 \ \text{kpc} \lesssim r \lesssim 15$~kpc and that in the rest of the disc (see Figure~\ref{fig:UGC7899_ellipse_gap}). It is also the only HIghMass galaxy imaged in \Hone thus far that shows evidence for an outer warp in that tracer \citep{hallenbeck2016highmass}. Among galaxies examined in this work, then, UGC~7899 is the best candidate for late accretion, where both hot-mode (outer disc radial flows; e.g. \citealt{pezzulli2016accretion}) and cold-mode (outer disc warp; e.g. \citealt{shen2006galactic}) accretion mechanisms may be at work. Attributing these features to late accretion is premature, however, given the well-known degeneracy between disc geometry and non-circular flows in kinematic models (see Section~\ref{subsec:kinematicmodels}). If UGC~7899 does harbour a warp that begins far within the optical disc (Section~\ref{subsubsec:7899}), then the application of flat-disc models in Section~\ref{subsec:ncflowsalldisk} may return spurious radial flow signatures. A cursory examination of this effect in the \Hone mapping data for UGC~7899 was carried out in \cite{hallenbeck2016highmass}, and a joint \Hone + \Halpha kinematic analysis may help disentangle a warp and any outer radial flows. We leave that significant task for future work.

Beyond the non-circular flows examined here, a future step for constraining the origin of the large \Hone reservoirs in HIghMass galaxies is to compare gas-phase metallicities to predictions from accretion models as well as to search for pristine accreting gas \citep{pezzulli2016accretion,fraternali2018angular}. SITELLE's narrow SN3 filter precludes this study from the data in-hand, but observations with its SN1 and SN2 filters would afford line ratio measurements to carry out such an investigation (e.g. \citealt{rousseau2018ngc628,moumen20193d,flagey2020wide}). But in this work, we have not found direct evidence for late accretion in the HIghMass galaxies examined.

\vspace{-8.5pt}
\section{Conclusion}

We have presented new \Halpha velocity maps of the HIghMass systems UGC~7899, UGC~8475, UGC~9037, and UGC~9334, obtained with the SITELLE Imaging Fourier Transform Spectrometer on CFHT, in order to search for non-circular flows indicating late gas accretion that might explain the large \Hone reservoirs in these galaxies.

The SITELLE hyperspectral velocity cubes were calibrated with the ORCS package to produce arcsecond velocity and velocity-uncertainty maps (Section~\ref{sec:observations_datareduction}). The maps for UGC~7899, UGC~9037 and UGC~9334 are suitable for detailed kinematic modelling, but the poor sensitivity of the UGC~8475 maps due to distributed photon noise from the Moon precludes searches for non-circular flows in this object.

We proceed to apply kinematic models to the velocity maps for UGC~7899, UGC~9037 and UGC~9334 using the DiskFit algorithm (Section~\ref{sec:dataanalysis}). We find no evidence for radial flows in the outer discs of UGC~9037 and UGC~9334, placing conservative upper limits of $\Vr \lesssim 15 \ \kms$ and $\Vr \lesssim 20 \ \kms$ for $r \gtrsim 5$~kpc, respectively (Section~\ref{subsec:ncflowsalldisk}). There are hints of outer disc radial flows in UGC~7899 with amplitudes as large at $\Vr \sim 50 \ \kms$.

We find clear signatures of non-circular flows in the inner discs of UGC~7899, UGC~9037 and UGC~9334. The major axis alignment of a bar-like optical feature in UGC~9334 precludes detailed modelling of that system, but we apply rotation-only, inner radial flow models and inner bisymmetric models to the velocity maps for UGC~7899 and UGC~9037 to determine the likely origin of those asymmetries (Section~\ref{subsec:appofkinmodels}). We find that inner radial flow and inner bisymmetric models describe their kinematics equally well, and determine our preferred model for each galaxy by carrying out disc-bulge-bar decompositions of \textit{R}-band imaging using the photometric branch of DiskFit and requiring physical consistency with the kinematic models (Section 3.4). As a result of this exercise, our preferred model for UGC~7899 is found to be a disc with inner radial flows and marginal non-circular flows beyond, whereas UGC~9037 is best characterized as a disc with an inner bisymmetric flow and axisymmetric rotation in the outer disc.

Our preferred model for UGC~7899 is an axisymmetric outer disc with inner radial inflows. In this scenario, its enhanced central \Hone surface density indicates that UGC~7899 may be in the midst of a significant star formation episode (Section~\ref{subsec:compwithlit}). We do not corroborate the coherent outer disc flows in this system implied by some lower-resolution \Hone kinematic models, but there are hints of asymmetries at large $r$ in our \Halpha maps including non-zero radial flow components, asymmetries in the \Halpha distribution as well as a dip in the rotation curve that likely stems from a disconnect between the gas in the Southern half of the disc at $10 < r < 15$~kpc and that in the rest of the disc.

We posit that UGC~9037 harbours a weak inner bar that is shuttling gas toward the galaxy centre, consistent with the high inner \Hone and CO surface density measured in this system and suggesting an upcoming episode of star formation (Section~\ref{subsec:compwithlit}). While we only favour a bar model for this galaxy to explain its inner-disc kinematics, the photometric and kinematic signatures of weak bars in UGC~7899 and UGC~9334 as well suggest that, as with the broader galaxy population, these structures may be common in HIghMass galaxies.

With hints of outer disc radial flows and an outer disc warp, UGC~7899 is the best candidate for late accretion among the galaxies examined (Section~\ref{subsec:radialflowsandlateaccretion}). However, this interpretation is muddled by the well-known degeneracy between non-circular flows and disc warps in kinematic models, which may be at work in the flat-disc DiskFit models that we have applied. A careful joint \Hone + \Halpha kinematic analysis of UGC~7899 that allows for outer disc warps may help disentangle these effects, and gas-phase metallicity information would also add additional constraints on late accretion. 

In this work, we have found weak constraints on hot-mode accretion from upper limits on outer disc radial flows implied by our data, but have not found direct evidence for accretion in the HIghMass galaxies examined.

\section*{Acknowledgements}

We thank Dr.\ Thomas Martin and Dr.\ Laurent Drissen whose efforts to develop ORCS were instrumental. KS acknowledges support from the Natural Sciences and Engineering Research Council of Canada (NSERC).

The observations in this work were obtained at the Canada-France-Hawaii Telescope (CFHT) which is operated from the summit of Maunakea by the National Research Council of Canada, the Institut National des Sciences de l'Univers of the Centre National de la Recherche Scientifique of France, and the University of Hawaii. The observations at the Canada-France-Hawaii Telescope were performed with care and respect from the summit of Maunakea which is a significant cultural and historic site. 

We thank our anonymous reviewer, whose comments have greatly improved this paper.


\section*{Data availability}

The data underlying this article will be shared on reasonable request by the corresponding author.

\bibliographystyle{mnras}
\bibliography{Bisaria_et_al_HIghMass} 


\appendix

\bsp	
\label{lastpage}
\end{document}